%% file: ANA-EXOT-2020-01-PAPER.tex
\pdfoutput=1
\documentclass[PAPER, cernpreprint, atlasdraft=false, texlive=2023, USenglish, texmf, orcidlogo]{atlasdoc}
\usepackage{atlaspackage}
\usepackage{atlasxref}
\usepackage{atlasbiblatex}

\usepackage{atlasphysics}
\usepackage{slashed}
\usepackage{feynmf}
\usepackage{graphicx}
\usepackage{placeins}
\usepackage{multirow}
\graphicspath{{logos/}{figures/}}

\usepackage{ANA-EXOT-2020-01-PAPER-defs}

\AtlasTitle{Search for singly produced vector-like top partners  in multilepton final states with $139$~{\mbox{fb\(^{-1}\)}}~of $pp$ collision data at $\sqrt{s} = 13$~TeV with the ATLAS detector}

\AtlasVersion{2.4}

\AtlasAbstract{%
A search for the single production of a vector-like top partner ($T$) with mass greater than 1~TeV decaying into a $Z$ boson and a top quark is presented, using the full Run 2 dataset corresponding to 139~{\mbox{fb\(^{-1}\)}}~of $pp$ collisions at $\sqrt{s} = 13$~TeV, collected in 2015–2018 with the ATLAS detector at the Large Hadron
Collider.
The targeted final state is characterized by the presence of a pair of electrons or muons with opposite-sign charges which form a $Z$-boson candidate, as well as
by the presence of $b$-tagged jets and forward jets.
Events with exactly two or at least three leptons are categorized into two independently optimized analysis channels.
No significant excess above the background expectation is observed and the results from the two channels are statistically combined to set exclusion limits at 95\% confidence level on the masses and couplings of $T$. The results are interpreted in several benchmark scenarios to set limits on the mass and universal coupling strength ($\kappa$) of the vector-like quark. For singlet $T$ quarks, $\kappa$ values between 0.22 and 0.64 are excluded for masses between 1000 and 1975~GeV. For $T$ quarks in the doublet scenario, where the production cross section is much lower, $\kappa$ values between 0.54 and 0.88 are excluded for masses between 1000 and 1425~GeV.
}

\AtlasRefCode{EXOT-2020-01}
\PreprintIdNumber{CERN-EP-2023-074}

\AtlasJournal{Phys.\ Rev.\ D}
\AtlasJournalRef{Phys.\ Rev.\ D 109 (2024) 112012}
\AtlasDOI{10.1103/PhysRevD.109.112012}

\hypersetup{pdftitle={ATLAS document},pdfauthor={The ATLAS Collaboration}}

\begin{document}

\maketitle

\section{Introduction}
\label{sec:intro}

The discovery of the Higgs boson by the ATLAS and CMS experiments at the Large Hadron Collider (LHC)~\cite{HIGG-2012-27,CMS-HIG-12-028} was a
major milestone for the Standard Model (SM), the framework for understanding the fundamental constituents of our universe and their interactions. While finding the
Higgs boson completed the SM, this model leaves  many  questions  unaddressed about the physical nature of our universe.
One of these questions is
about
the measured  Higgs boson mass--specifically, the many orders of magnitude that separate it from the
Planck scale, requiring quadratically divergent finely tuned corrections~\cite{Naturalness}. These large corrections are mitigated in some beyond Standard Model (BSM) theories.
Vector-like Quarks (VLQs), colored spin-$\frac{1}{2}$ fermions with identical electroweak (EW) representation for both left and right chiralities, are predicted by a
large subset of these theories~\cite{new-fermion, comp-higgs, comp-higgs-2, comp-higgs-3, littlest-higgs, extra-dim, extra-dim-2, extra-dim-bounds, VL-fermion, TP-LHC}.
The usual four VLQ species are denoted by $X_{+\frac{5}{3}}, T_{+\frac{2}{3}}, B_{-\frac{1}{3}}$ and $Y_{-\frac{4}{3}}$, where the subscript indicates
the electric charge of the corresponding particle. Their renormalizable EW representation consists of $(T)$ or $(B)$ singlets, $(X,T)$, $(T,B)$, or $(B,Y)$ doublets
and $(X,T,B)$ or $(T, B, Y)$ triplets. In most representations, they couple to the SM quarks via an exchange of charged $(W^\pm)$ or neutral $(Z, H)$ bosons.
The interaction of the VLQs with the SM quarks can be summarized with the following simplified Lagrangian~\cite{Fuks}:

\begin{equation}\label{eqn:SimpLag}
\mathcal{L} = \sum_{Q, q, \zeta} \left[         \frac{g_w}{2c_W}     \tilde{\kappa}_\zeta^{Qq} \bar{Q}\slashed{Z} P_\zeta q
+       \frac{g_w}{\sqrt{2}}{\kappa}_\zeta^{Qq}       \bar{Q}\slashed{W} P_\zeta q
-                            \hat{\kappa}_\zeta^{Qq} H \bar{Q} P_\zeta q\right] + \mathrm{h.c.,}
\end{equation}

where $Q$ represents a VLQ, $\zeta$  represents the chirality with $P_\zeta$ being the corresponding projection operator, $q$ represents a SM quark of up or down type, and the
electroweak couplings $\tilde{\kappa}_\zeta^{Qq},  {\kappa}_\zeta^{Qq},$ and $\hat{\kappa}_\zeta^{Qq}$  determine
the coupling strengths between $Q$ and $q$ when mediated by the $Z,W,$ and $H$ bosons respectively.
The mass hierarchy of the SM quarks suggests that VLQs interact predominantly with the third-generation SM quarks~\cite{okada,Aguilar}.
Hence, VLQ interactions with lighter generations are set to zero in the simplified representation of Eq.~(\ref{eqn:SimpLag}) and the VLQ species $T$ and $B$ are referred to as top and bottom partners respectively  in the following discussion.

At the LHC, the dominant modes of VLQ production are pair and single production. During Run~1 (2010--2012) of the LHC, the VLQ searches at center-of-mass energies of $\sqrt{s} = $ 7 and 8~\TeV  typically probed VLQ masses lower than 1~\TeV where VLQ pair production dominates~\cite{Aguilar}.
Since the pair production of VLQs proceeds through QCD interactions, the cross-section under the narrow-width approximation only depends on the VLQ mass and not on the species nor its EW representation.
No significant excess was observed by the Run~1 searches looking for pair production of VLQs  and lower exclusion limits on VLQ mass were reported in the range of approximately \mbox{600--1000 GeV}, depending on the search~\cite{EXOT-2013-18, EXOT-2014-17, EXOT-2013-17, CMS-B2G-13-005, CMS-B2G-13-006, CMS-B2G-12-012}.
The searches for single production of  $T$ or $B$ also reported similar limits in Run~1 for a range of couplings probed in the singlet and doublet benchmarks~\cite{EXOT-2013-17, EXOT-2014-12}.

Run~2 (2015--2018) searches at $\sqrt{s} = $~13~TeV for pair production of top and bottom partners from ATLAS and CMS have set lower limits on the VLQ masses in the range of $800 - 1400$~GeV~\cite{EXOT-2016-13, EXOT-2016-14, EXOT-2016-15,  EXOT-2016-16, EXOT-2017-14, EXOT-2017-17, ATLAS-TZt-PP-22, CMS-B2G-16-024, CMS-B2G-17-003, CMS-B2G-17-011, CMS-lept-PP-22}.
Since the single production of VLQs can have a larger cross-section at high masses for reasonably large couplings~\cite{Aguilar}, the Run~2 searches are increasingly focusing on this production mode~\cite{EXOT-2016-35, EXOT-2016-17, EXOT-2017-16, CMS-B2G-17-007,  CMS-B2G-17-018, CMS-B2G-17-009, CMS-B2G-18-003, CMS-SP-TMET-2022, CMS-THt-SP-23, ATLAS-allhad-run2full, ATLAS-TZtHt-SP-23}.
However, unlike pair production, single production of VLQs is mediated by electroweak processes.
Both the kinematics and cross-sections of the single VLQ processes depend on the representation and the choice of couplings that determine the relative strength of these particles
interacting with SM quarks and vector/Higgs bosons.

This \paper focuses on the search for a singly produced $T$ decaying to a $Z$ boson and a top quark using the complete Run~2 dataset collected by the ATLAS detector. As shown in Figure~\ref{fig:feyn-WTZt-ZTZt}, the electroweak interaction can produce a $T$
in the $s$- and $t$-channel topologies.
These processes are referred to as $WTZt$ and $ZTZt$ processes, where the first $W$ or $Z$ refers to the vector boson scattered off of the incoming parton, $T$ refers to the VLQ exchanged in the intermediate stage, and $Zt$ refers to the production of the $Z$ boson and $t$ quark. The $WTZt$ mode dominates in the singlet representation of $T$ since the smaller coupling between the $T$ and the $Z$ boson and the kinematically disfavored $g\rightarrow t\bar{t}$ splitting suppress the contribution from $ZTZt$. On the other hand, the latter is the dominant mode in the doublet representation since the $T$'s coupling with the $W$ boson is suppressed by the mass diagonalization matrix~\cite{Aguilar}.
A model-independent approach~\cite{Atre,Panizzi,Wulzer} that allows probing a wide range of couplings for the $W,Z,$ and $H$-mediated interactions of these heavy quarks  is adopted in this search.

The analysis is performed across two channels spanning decays of $T$ quarks with mass in the range between 1.0 and 2.7~TeV.
The dilepton (\dilep) channel  selects events with exactly one pair of opposite-sign leptons, i.e. electrons or muons and a hadronically decaying top quark in the final state. The trilepton (\trilep)  channel requires,  in addition to the pair of opposite-sign leptons, at least a third lepton from a leptonically decaying top quark.
The reconstructed $Z$-boson from the pair of leptons is expected to have large transverse momentum while the hadronic decay of the boosted top quark in the \dilep channel would constitute wide-radius jet signatures.
These results improve on a previous ATLAS search~\cite{EXOT-2016-35} with a data sample corresponding to an integrated luminosity of 36.1~\ifb, by benefiting from the larger data sample of 139~\ifb, improved kinematic selections and introducing a method to identify hadronically decaying top quarks.

The \paper is organized as follows.
Section~\ref{sec:detector} provides a brief overview of the ATLAS detector.
The data  and simulated  background and signal samples are described in Section~\ref{sec:samples}.
Section~\ref{sec:reco} describes the object definition while event selection, categorization, and the analysis strategy are outlined in Section~\ref{sec:selection}.
The systematic uncertainties are discussed in Section~\ref{sec:syst}. Results obtained from this search are reported in Section~\ref{sec:stat-result} and conclusions are presented in Section~\ref{sec:conclusion}.

\begin{figure*}[!ht]
\centering
\subfloat[]{
\includegraphics[width=0.35\textwidth]{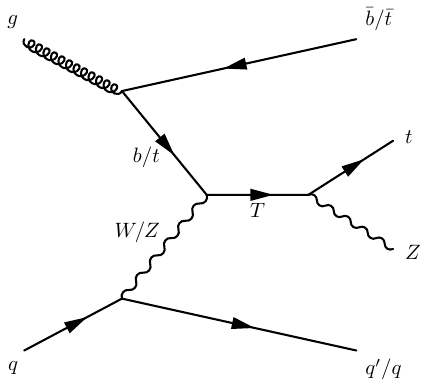}
\label{fig:feyn-WZTZt-s}
}
\subfloat[]{
\includegraphics[width=0.35\textwidth]{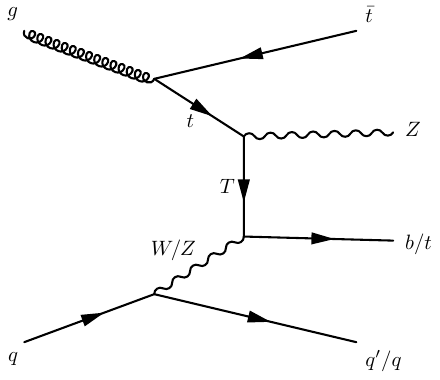}
\label{fig:feyn-WZTZt-t}
}
\label{fig:feyns-2}
\caption{ Illustrative \protect\subref{fig:feyn-WZTZt-s} $s$-channel and \protect\subref{fig:feyn-WZTZt-t} $t$-channel Feynman diagrams for $WTZt$ and $ZTZt$ processes.
}
\label{fig:feyn-WTZt-ZTZt}
\end{figure*}


%

%
\section{ATLAS detector}
\label{sec:detector}

%
%

\newcommand{\AtlasCoordFootnote}{%
ATLAS uses a right-handed coordinate system with its origin at the nominal interaction point (IP)
in the center of the detector and the \(z\)-axis along the beam pipe.
The \(x\)-axis points from the IP to the center of the LHC ring,
while the \(y\)-axis points upwards.
Cylindrical coordinates \((r,\phi)\) are used in the transverse plane,
\(\phi\) being the azimuthal angle around the \(z\)-axis.
The pseudorapidity is defined in terms of the polar angle \(\theta\) as \(\eta = -\ln \tan(\theta/2)\).
Angular distance \((\Delta R)\) is defined as \(\Delta R \equiv \sqrt{(\Delta\eta)^{2} + (\Delta\phi)^{2}}\).}

The ATLAS detector~\cite{PERF-2007-01} at the LHC covers nearly the entire solid angle around the collision point.\footnote{\AtlasCoordFootnote}
It consists of an inner tracking detector surrounded by a thin superconducting solenoid, electromagnetic and hadronic calorimeters,
and a muon spectrometer incorporating three large superconducting air-core toroidal magnets.

The inner-detector system (ID) is immersed in a \SI{2}{\tesla} axial magnetic field
and provides charged-particle tracking in the range of \(|\eta| < 2.5\).
The high-granularity silicon pixel detector covers the vertex region and typically provides four measurements per track,
the first hit normally being in the insertable B-layer (IBL) installed before
Run~2~\cite{PIX-2018-001}.
It is followed by the silicon microstrip tracker (SCT), which usually provides eight measurements per track.
These silicon detectors are complemented by the transition radiation tracker (TRT),
which enables radially extended track reconstruction up to \(|\eta| = 2.0\).
The TRT also provides electron identification information
based on the fraction of hits (typically 30 in total) above a higher energy-deposit threshold corresponding to transition radiation.

The calorimeter system covers the pseudorapidity range  of \(|\eta| < 4.9\).
Within the region \(|\eta|= 3.2\), electromagnetic calorimetry is provided by barrel and
endcap high-granularity lead/liquid-argon (LAr) calorimeters,
with an additional thin LAr presampler covering \(|\eta| < 1.8\)
to correct for energy loss in material upstream of the calorimeters.
Hadron calorimetry is provided by the steel/scintillator-tile calorimeter,
segmented into three barrel structures within \(|\eta| = 1.7\), and two copper/LAr hadron endcap calorimeters.
The solid angle coverage is completed with forward copper/LAr and tungsten/LAr calorimeter modules
optimised for electromagnetic and hadronic energy measurements respectively.

The muon spectrometer (MS) comprises separate trigger and
high-precision tracking chambers measuring the deflection of muons in a magnetic field generated by the superconducting air-core toroidal magnets.
The field integral of the toroids ranges between \num{2.0} and \SI{6.0}{\tesla\metre}
across most of the detector.
Three layers of precision chambers, each consisting of layers of monitored drift tubes, cover the region \(|\eta| < 2.7\),
complemented by cathode-strip chambers in the forward region, where the background is highest.
The muon trigger system covers the range of \(|\eta| < 2.4\) with resistive-plate chambers in the barrel, and thin-gap chambers in the endcap regions.

Interesting events are selected by the first-level trigger system implemented in custom hardware,
followed by selections made by algorithms implemented in the software-based high-level trigger (HLT)~\cite{TRIG-2016-01}.
The first-level trigger reduces the incoming data rate from the \SI{40}{\MHz} bunch crossings to a design value of  \SI{100}{\kHz}, which is further reduced by the HLT  to record events to disk at a rate of about \SI{1}{\kHz}.

An extensive software suite~\cite{ATL-SOFT-PUB-2021-001} is used in data simulation, in the reconstruction and analysis of real
and simulated data, in detector operations, and in the trigger and data acquisition systems of the experiment.


%

%
\section{Data and simulated event samples}
\label{sec:samples}

%
%
%
%

The  data sample was collected by the ATLAS detector in proton--proton $(pp)$ collisions
at a center-of-mass energy of $\sqrt{s}=\SI{13}{\TeV}$  during Run~2 with all detector
subsystems operational and with the LHC operating in
stable beam conditions with $\SI{25}{ns}$ bunch spacing.
The dataset corresponds to an integrated luminosity of 139~\ifb~with an average of about 34 simultaneous interactions per bunch crossing (pileup).

All of the nominal Monte Carlo (MC) simulation samples used in the analysis were processed with the ATLAS
simulation framework~\cite{SOFT-2010-01}, using a detailed simulation based on \GEANT~\cite{Agostinelli:2002hh}.
The effect of \pileup was modeled by overlaying the simulated hard-scattering event with inelastic $pp$ events generated with \PYTHIA[8.186]~\cite{Sjostrand:2007gs} using the \NNPDF[2.3lo] set of parton distribution functions (PDF)~\cite{Ball:2012cx} and the A3 set of tuned parameters (tune)~\cite{ATL-PHYS-PUB-2016-017}. The \pileup is reweighted in the MC event samples to match the pileup conditions in data.

The nominal MC sample for $Z$ boson production  with jets (\zjets) was generated with
\SHERPA[2.2.1]~\cite{Bothmann:2019yzt}
and the nominal diboson (\VV) sample
was generated with \SHERPA[2.2.2]. Both  the samples were generated using the \NNPDF[3.0]~\cite{Ball:2014uwa} next-to-next-to-leading-order (NNLO) PDF set.
The \zjets sample includes events generated with up to two partons at  next-to-leading-order
(NLO) and up to four partons at leading-order (LO) and is normalized to the NNLO cross-section~\cite{Anastasiou:2003ds}.
The  \VV sample is normalized to the \SHERPA\ NLO cross-section and includes $q\bar{q}$-initiated events with up to one parton
at NLO and up to three partons at LO and $gg$-initiated processes simulated using LO matrix elements for up to one additional jet.
For both samples, \textsc{Comix}~\cite{Gleisberg:2008fv} and \OPENLOOPS~\cite{Buccioni:2019sur,Cascioli:2011va,Denner:2016kdg} were used and
the matrix element (ME) was merged with the \SHERPA\ parton shower~\cite{Schumann:2007mg} according to the \MEPSatNLO
prescription~\cite{Hoeche:2011fd,Hoeche:2012yf,Catani:2001cc,Hoeche:2009rj}.
To estimate modeling uncertainties for these backgrounds, additional samples were produced with
\MGNLO[2.2.3]~\cite{Alwall:2014hca}, using the \NNPDF[3.0nlo] PDF set and interfaced to
\PYTHIA[8.210]~\cite{Sjostrand:2014zea} with the A14 tune~\cite{ATL-PHYS-PUB-2014-021} and the \NNPDF[2.3lo] PDF for showering.
In the \dilep channel, the simulated $Z$+jets events are categorized into $Z$+Light Flavor (LF) and $Z$+Heavy Flavor (HF) events in accordance with the
absence or presence of heavy flavor jets in the MC generator record of these simulated events in the kinematic acceptance defined in Sections~\ref{sec:reco}~and~\ref{sec:selection}.
An additional set of \zjets events was generated with \SHERPA[2.2.11] to compare the modeling of the \zjets background in the \dilep channel.

The nominal \ttbar~background sample uses the \POWHEG
method~\cite{Nason:2004rx, Frixione:2007vw}
implemented in \POWHEGBOX~v2~\cite{Alioli:2010xd, Campbell:2014kua} with the \NNPDF[3.0nnlo] PDF
set. \POWHEGBOX was interfaced with \PYTHIA[8.230]
with the A14 tune for showering and hadronization.
The sample is normalized to the NNLO cross-section in QCD including resummation
of next-to-next-to-leading-logarithmic (NNLL) soft gluon
terms calculated with \TOPpp~\cite{Czakon:2011xx,Czakon:2013goa, Beneke:2011mq, Cacciari:2011hy, Czakon:2012pz, Czakon:2012zr, Baernreuther:2012ws}.
For the evaluation of modeling uncertainties, samples were produced with the same matrix element generator as the nominal
sample, but \HERWIG[7] was used with the H7-UE-MMHT tune~\cite{Bellm:2015jjp} for showering and hadronization.
To assess the uncertainty in the matching of NLO matrix elements to the
parton shower, additional samples were generated with
\MGNLO[2.6.0] and the \NNPDF[3.0nnlo] PDF set, interfaced with \PYTHIA[8.230] using the same showering configuration as the nominal sample.

The nominal sample for \ttbar production  with a vector boson (\ttbarW and \ttbarZ) was generated  with \MGNLO[2.3.3] interfaced with \PYTHIA[8.210] for
showering and hadronization, using the \NNPDF[2.3lo] PDF set and the A14 tune.
These events were normalized to the  NLO cross-sections calculated with \MGNLO~\cite{frederix2018large, frixione2015electroweak}.
To evaluate modeling uncertainties, alternative samples were produced
where either the A14 tune was varied or \HERWIG[7] was used with the H7-UE-MMHT tune for the showering.  Additional samples were produced using \SHERPA[2.2.1] to evaluate the uncertainty due to the choice of matrix element generator.
The production of $t\bar{t}t\bar{t}$ and $t\bar{t}WW$ events was modeled using the \MGNLO generator at LO with the \NNPDF[3.1nlo] PDF, interfaced with \PYTHIA[8.230] using the A14 tune and the \NNPDF[2.3lo] PDF set. These samples were normalized to cross-sections calculated with NLO QCD and EW corrections~\cite{frederix2018large}. These four processes are merged into a single process labeled \ttbarV  in which the contribution from \ttbarZ dominates due to the requirement of an opposite-sign same-flavor (OS-SF) lepton pair with an invariant mass close to that of a $Z$ boson in the final state.

The single-top-quark processes were simulated with \POWHEGBOX~\cite{Frederix:2012dh,Re:2010bp}
using the \NNPDF[3.0nlo] PDF set and interfaced to \PYTHIA[8.234] with the A14 tune. The
samples are normalized to their respective NLO QCD cross-sections~\cite{Aliev:2010zk, Kant:2014oha} for the $t$-channel
and $s$-channel, and with additional
NNLL soft gluon terms for $Wt$ production~\cite{Kidonakis:2011wy, Kidonakis:2010ux,Kidonakis:2010tc}.
The production of $tZq$ and $tWZ$ events was modeled using the \MGNLO
generator at NLO with the \NNPDF[3.0nlo] PDF and interfaced with \PYTHIA[8.212] using the A14 tune and the \NNPDF[2.3lo] PDF set.  The diagram-removal scheme~\cite{Frixione:2008yi} was used in the generation of $Wt$ and $tWZ$ events to address their overlaps with \ttbar and \ttbarZ samples. These samples were also merged into a single process labeled as single-top.

Signal samples for the $WTZt$ and $ZTZt$ processes were generated at LO using \MGNLO with the Universal FeynRules Output (UFO) model~\cite{UFO} from Ref.~\cite{Fuks}, interfaced with \PYTHIA[8.244] and using the \NNPDF[3.0lo] PDF set and the A14 tune. This model uses the four-flavor scheme.
Top partner samples were obtained for various masses, ranging from  1.0~\TeV and 2.7~\TeV in steps of 100~\GeV.
For each mass point, the coupling strength $\kappa$ was varied between 0.1 and 1.0 with a step of 0.05 (0.1) for couplings less (greater) than 0.5.
The relative strength of the three coupling modes were obtained from $\vec{\xi} = (\xi_W, \xi_Z, \xi_H) = (0.5, 0.25, 0.25)$.\footnote{The parameterization of VLQ Lagrangian in terms of the $\kappa$ and $\vec{\xi}$ parameters was introduced in Ref.~\cite{Panizzi} and is used for the interpretation of the search presented in this \paper. Its conversion to the coupling convention in Eq.~(\ref{eqn:SimpLag}) is obtained by doing a one-to-one mapping of the tree level couplings in the Lagrangian. The $\xi_W, \xi_Z, \xi_H$ parameters satisfy the constraint  $\xi_W + \xi_Z + \xi_H = 1$ and asymptotically represent the branching ratios of $T$ decaying into $Wb,Zt,$ and $Ht$  in the narrow-width limit.}
The MC samples were produced for a subset of the mass and coupling points. Those samples were used for other masses and couplings by implementing an event-by-event matrix element reweighting~\cite{reweighting}.
While the left- and right-handed couplings dominate in the singlet and doublet representations of $T$, the current analysis was found to be insensitive to the choice of chirality for the signal MC generation and all signal samples were generated using the left-handed couplings.
The signal samples were normalized to the NLO cross-section calculated with the narrow-width approximation~\cite{Panizzi-NLO}. Due to the requirement of at least one jet containing a $b$-hadron in the signal region the kinematic differences between the 4 flavor scheme and 5 flavor scheme are expected to be small. However, since the top partner can have large width based on the choice of coupling, additional correction factors were applied to evaluate
the cross-section at finite width~\cite{interpretation}. The contribution from the non-resonant $t$-channel diagram was also taken into account~\cite{vlqnonres}.


%

%
\section{Object reconstruction}
\label{sec:reco}

%
%
%
%

%
%

Events are required to have at least one vertex  with at least
two tracks with transverse momentum $\pt > \SI{0.5}{\GeV}$. The primary vertex (PV)
is defined as the one with the largest $\Sigma \pt^2$, where the sum is performed over all the tracks matched to the vertex.

Electrons are reconstructed from energy clusters in the EM calorimeter matched with ID tracks
and must fulfill the \textit{tight likelihood}
identification criteria~\cite{EGAM-2018-01}. Electrons are
calibrated~\cite{EGAM-2018-01} and  required to have
$\pt > \SI{28}{\GeV}$ and to be reconstructed  within $|\eta|=2.47$,
excluding the calorimeter barrel--endcap transition regions ($1.37 < |\eta|< 1.52$).
A high acceptance for the expected signal events is maintained by not applying an isolation requirement to electron candidates
beyond those implicit in the trigger requirements, which are explained in Section~\ref{sec:selection}. Furthermore, the track matched with the candidate electron is required to have a longitudinal impact parameter with
respect to the PV which satisfies $|z_0 \cdot \sin\theta| < \SI{0.5}{mm}$ and a transverse impact parameter with
respect to the beamline ($d_0$) with a significance $|d_0|/\sigma(d_0) < 5$.

Muons are
reconstructed~\cite{MUON-2018-03} from combined tracks in the MS
and the ID and must fulfill the \textit{medium}
identification criteria~\cite{MUON-2018-03}. Muons are calibrated and required to have $\pt>\SI{28}{\GeV}$ and to be
reconstructed  within $|\eta|=2.5$.  Muon candidates must
also satisfy the track-based isolation requirements
defined by the \textit{FixedCutTightTrackOnly} working point~\cite{MUON-2018-03}.
This isolation requirement does not significantly reduce the acceptance.
This working point uses the scalar sum of the \pt of all tracks that are inside of a cone of size
$\Delta R = \min\left\{0.3,\SI{10}{\GeV}/\pt(\mu)\right\}$
around the muon candidate,
where $\pt(\mu)$ is the candidate muon $\pt$. The track matched with the muon candidate under consideration is excluded from the sum.
The muon is selected if this sum is less than 15\% of $\pt(\mu)$.
Finally, muon candidates are required to
have a matched track with $|z_0 \cdot \sin\theta| < \SI{0.5}{mm}$
and a significance $|d_0|/\sigma(d_0) < 3$.

Jets are reconstructed from particle-flow objects~\cite{pflow} using the \antikt\ algorithm~\cite{Cacciari:2008gp, Fastjet}
with a radius parameter of $R = 0.4$.
Jets are calibrated to the particle level from a combination
of simulation-based corrections and measurements in data~\cite{jes}
and are required to fulfill $\pt > \SI{25}{\GeV}$ for
$|\eta|<2.5$ and $\pt>\SI{35}{\GeV}$ for $2.5<|\eta|<4.5$. These two jet categories are respectively called
central and forward jets. To reduce jet contributions from pileup, a `jet vertex tagger' algorithm
using a two-dimensional likelihood discriminant~\cite{PERF-2014-03}
is applied to jets with $|\eta|<2.4$ and $\pt < \SI{60}{\GeV}$.
The DL1r
algorithm~\cite{FTAG-2019-07} %
is used to
identify jets  in the central region ($|\eta|<2.5$)
containing $b$-hadrons ($b$-tagging) with a working point corresponding
to a $b$-tagging efficiency in simulated $\ttbar$ events of $77\%$,
a $c$-jet rejection factor of ${\sim}6$, and a light-jet rejection factor of ${\sim}192$.

The missing transverse
momentum~\cite{ATLAS-CONF-2018-023},%
with magnitude \met,
is defined as the negative vectorial
sum of the transverse momenta of all the calibrated reconstructed lepton and jet candidates in the event and includes a `soft term' with contributions
from tracks emanating from the PV but not matched with any of the reconstructed objects.

A procedure to remove potential overlaps between reconstructed lepton and jet candidates is
performed sequentially as follows.
First,
any electron sharing an ID track with one of the muons is removed. Next, any jet within $\Delta R =0.2$
of an electron is removed, followed by the removal of electrons within $\Delta R =0.4$ of
any remaining jet. Then
any jet with at most two tracks with $\pt>\SI{0.5}{\GeV}$ within $\Delta R= 0.2$ of a
muon is removed, unless it is $b$-tagged.
At the end of the procedure, any muon within
$\Delta R = \min\left\{0.4,0.04 + \SI{10}{\GeV}/\pt(\mu)\right\}$ of any remaining jet is removed.

To identify hadronically decaying boosted top-quark jets, variable-radius reclustered jets (vRC jets) are reconstructed using the variable-radius jet algorithm~\cite{krohn2009jets} from calibrated
jets
with $R = 0.4$
where the effective radius of the jet cone is chosen as $R_0 = \frac{2m_{t}}{p_T}$, $m_{t}$ and $p_T$ representing
the mass of the top quark and the jet transverse momentum respectively.
These vRC jets are known to provide stable top-tagging performance for a wide range of jet \pt~\cite{lapsien2016new}. The vRC jets are
called top-tagged (top-vetoed) if the vRC jet mass is greater (less) than 140~\GeV. Top-tagged and top-vetoed jets are
required to have \pt $>$ 200~\GeV and to contain at least two $R=0.4$ jets as jet constituents when their transverse momentum
is less than 700~\GeV.
%


%

%
\section{Event selection}
\label{sec:selection}

A common initial event selection is performed for both the \dilep and \trilep channels where events are required to have passed the single lepton trigger selections~\cite{TRIG-2016-01,TRIG-2018-01,TRIG-2018-05}.
For muons, triggers with a \pt threshold of 20~\GeV (26~\GeV) in 2015 (2016--2018) and isolation requirements, are combined in logical OR with triggers with a 50~\GeV \pt threshold with no isolation requirement.
A complementary barrel region only
trigger with a 60~\GeV \pt threshold is added for the 2017--2018 data taking period.
Similarly, electron triggers with isolation and identification requirements (tight likelihood identification in 2016--18, with a less restrictive requirement in 2015) and \pt thresholds of 24--26~\GeV
are combined with triggers with higher \pt thresholds
between 120~\GeV and 140~\GeV,
less restrictive identification criteria, and
no isolation requirements.

Events are additionally required to have at least one OS-SF lepton pair with the transverse momentum of each lepton being greater than 28~\GeV. The pair of OS-SF leptons with invariant mass \mll closest to the mass of the $Z$-boson (\mZ) is identified as the $Z$-boson candidate. Based on the expected hadronic activity of the signal topologies, events are also required to have  at least one vRC jet (in the \dilep~channel) or at least two central jets (in the \trilep~channel).

\subsection{Dilepton channel}
\label{subsec:SP2l}

%

The preselection of events for the \dilep channel (\dilepreg{PS}) requires the $Z$-boson candidate to have an invariant mass satisfying $|m(\ell\ell) - m_Z| < 10$~GeV.
Given that the search is sensitive to energetic final-state objects, additional background reduction is achieved by requiring the $Z$-boson candidate to have a transverse momentum (\ptz) of at least 200~\GeV and the scalar sum of transverse momenta of jets in the event, denoted by \HT, to be at least 300~\GeV.
Since signal events are expected to contain a hadronically decaying boosted top quark, events without any vRC jets are rejected.
The four momentum of the leading vRC jet $(J)$, defined as the vRC jet with the highest \pt, is combined with that of the $Z$-boson candidate to obtain an estimate of the mass of  $T$, $m_{\ell\ell J}$. To reduce sensitivity to VLQ pair-production and to allow for a consistent phenomenological interpretation of observed data counts, a requirement of $\HT + \met < m_{\ell\ell J}$ is also placed on preselection events~\cite{EXOT-2016-35}.

The main background for the \dilep channel is comprised of $Z$+jets events, with minor contributions from \diboson and \ttbar processes.
Measurements of differential cross sections have recently shown that the current simulations of $Z$+jets processes overestimate both the upper tail of the \HT spectrum and the cross section at high jet multiplicities~\cite{ATL-PHYS-PUB-2017-006}. This leads to discrepancies in the shapes of the \HT+\met and jet multiplicity spectra between the data and expected background in the kinematic regimes relevant for this search. Data-driven reweighting factors are therefore used to correct the observed discrepancies in these processes. These corrections are derived using the following procedure.

To improve the background modeling, a  reweighting factor is extracted for the $Z$+jets samples by comparing data with simulated events in the \dilepreg{PS} region for the jet multiplicity and $\HT + \met$ distributions.
These reweighting factors are separately derived for \dilepreg{PS} events with and without $b$-tagged jets. Since the shapes of the $Z$+LF and $Z$+HF samples were found to be similar in these regions, these factors are applied on
the $Z$+jets template obtained by combining these samples together.
The reweighting function is defined bin-by-bin by the formula

\begin{equation}
R_{\mathrm{Z+jets}}(x)=\frac{\mathrm{Data}(x)-\mathrm{MC}^{\mathrm{non-Z+jets}}(x)}{\mathrm{MC}^{\mathrm{Z+jets}}(x)},
\label{eqn:2l-reweighting}
\end{equation}
where $x$ represents the binning in each kinematic variable. After reweighting in jet multiplicity, a second reweighting is applied on $\HT + \met$ using Eq.~(\ref{eqn:2l-reweighting}).
The joint reweighting factors are applied on an event-by-event basis to estimate the MC contribution of $Z$+jets events for this channel. The
jet multiplicity
and $\HT + \met$ distributions in the \dilepreg{PS} region are shown in Figure~\ref{fig:2l_rwt}.
In the bins of jet multiplicity and $\HT + \met$ distributions, the contribution of non-$Z$-jets processes is less than 15\% before reweighting. The predicted signal contribution is also quite small. For a benchmark $WTZt$ process with $M_T = 1.5$~\TeV and $\kappa = 0.5$ in the singlet representation, the bin-by-bin signal-over-background ratio $(S/B)$ varies between 4--7\% for most bins and is always less than 9\%.

\begin{figure}
\centering
\subfloat[]{
\includegraphics[width=0.48\textwidth]{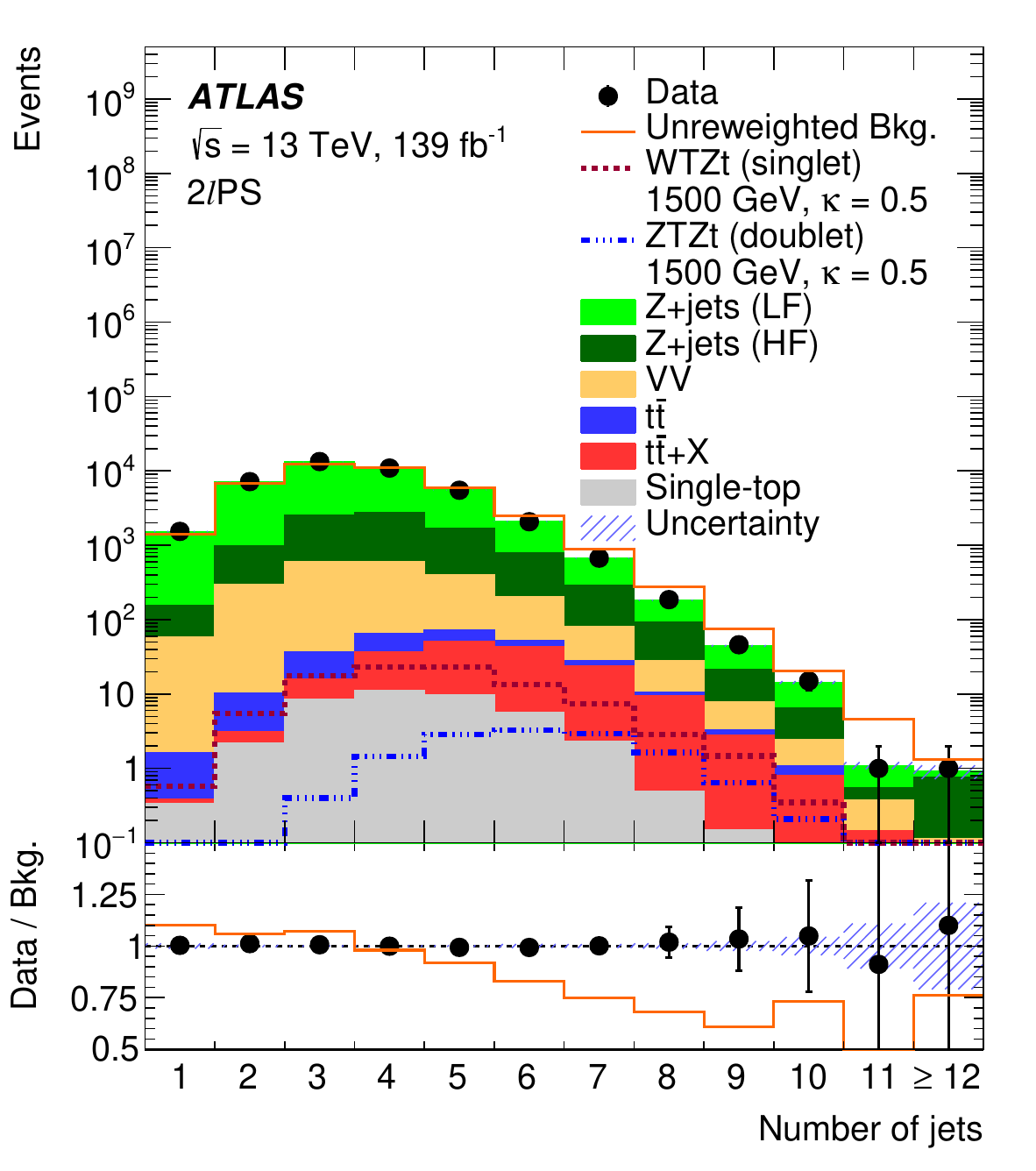}
\label{fig:2l_n_jet_ps}
}
\subfloat[]{
\includegraphics[width=0.48\textwidth]{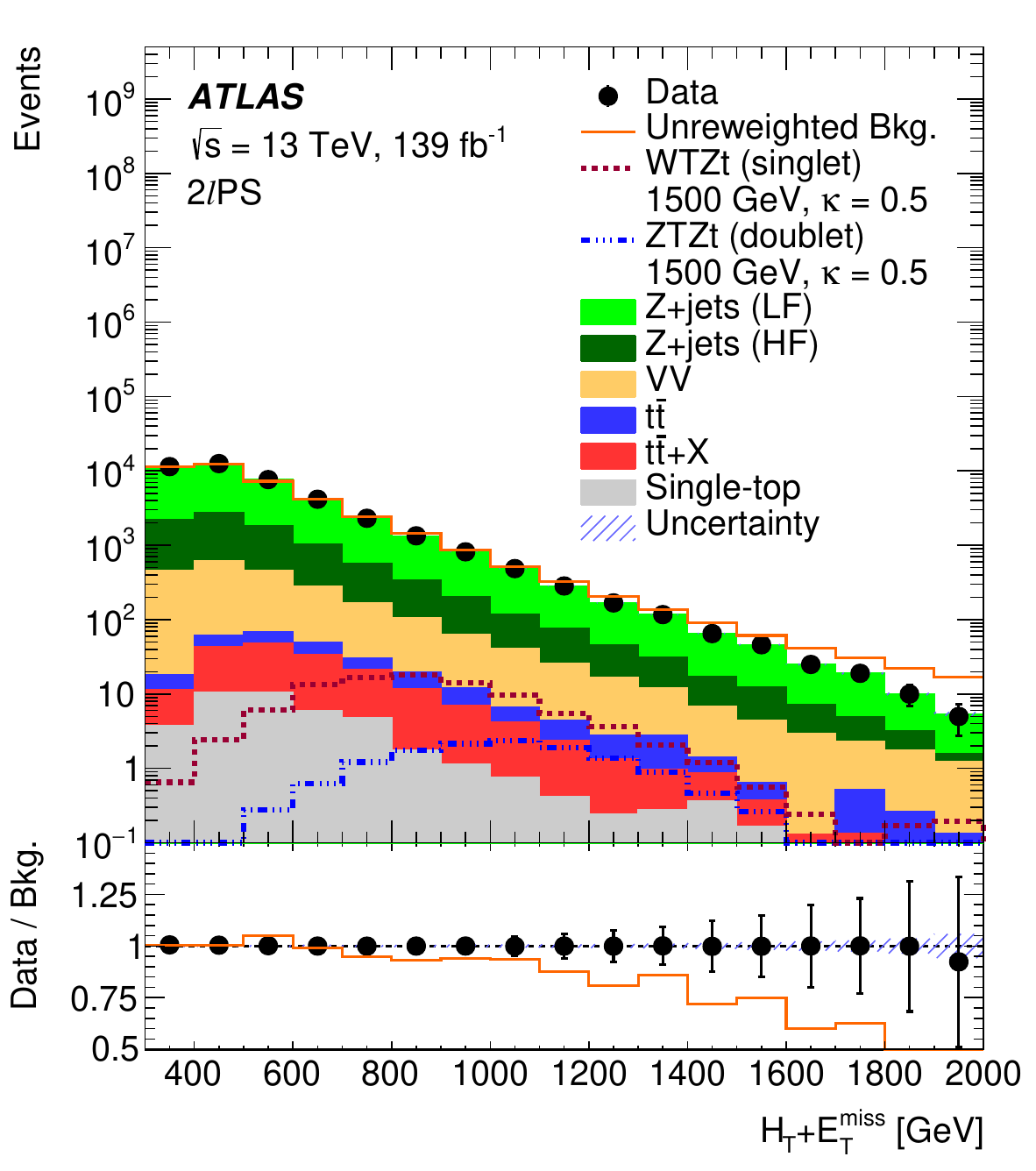}
\label{fig:2l_htmet_ps}
}
\caption[]{Distributions of \subref{fig:2l_n_jet_ps} jet multiplicity and \subref{fig:2l_htmet_ps} $\HT + \met$ in the \dilepreg{PS} region after reweighting. The overlaid solid line shows the total background prediction before the reweighting and the corresponding data-over-background ratio in the bottom panel. The red and blue dotted lines show the signal contributions from $WTZt$ and $ZTZt$ processes with $M_T = 1.5$~\TeV and $\kappa =  0.5$ in the singlet and doublet representations respectively. The last bin in each distribution contains the overflow.
}
\label{fig:2l_rwt}
\end{figure}

Figure~\ref{fig:2l_lineshapes} shows the distributions of the key kinematic variables for the expected background and benchmark signal processes in the \dilepreg{PS} region.
Events with a singly produced $T$ have a distinctive signature with a forward jet scattering off of a vector boson from one of the incoming partons.
Events are additionally expected to have $b$-tagged jets.
Hence, in addition to the preselection, the signal-enriched kinematic phase space, called the \textit{signal region} (\dilepreg{SR}), requires events with at least one
forward jet, at least one $b$-tagged jet, and at least one top-tagged jet, as shown in Table~\ref{tab:2lregions}.
The distribution of the transverse momentum of the $Z$-boson candidate, \ptz, was the variable that offered maximum sensitivity and, hence, is used as the final discriminant in this channel to perform the statistical fit described in Section~\ref{sec:stat-result}.

\begin{figure}
\centering
\subfloat[]{
\includegraphics[width=0.32\textwidth]{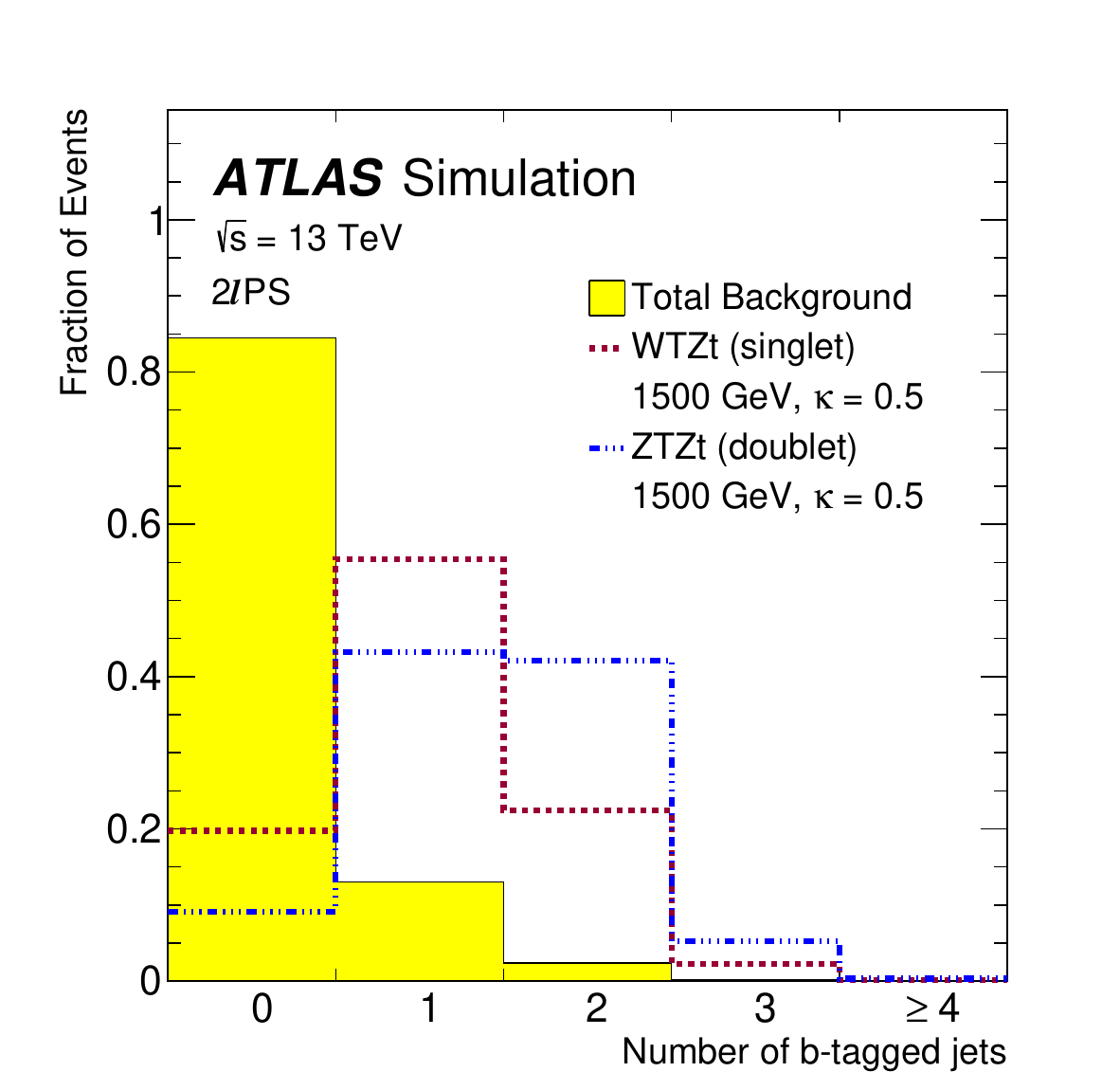}
\label{fig:2l_n_bjet_ps}
}
\subfloat[]{
\includegraphics[width=0.32\textwidth]{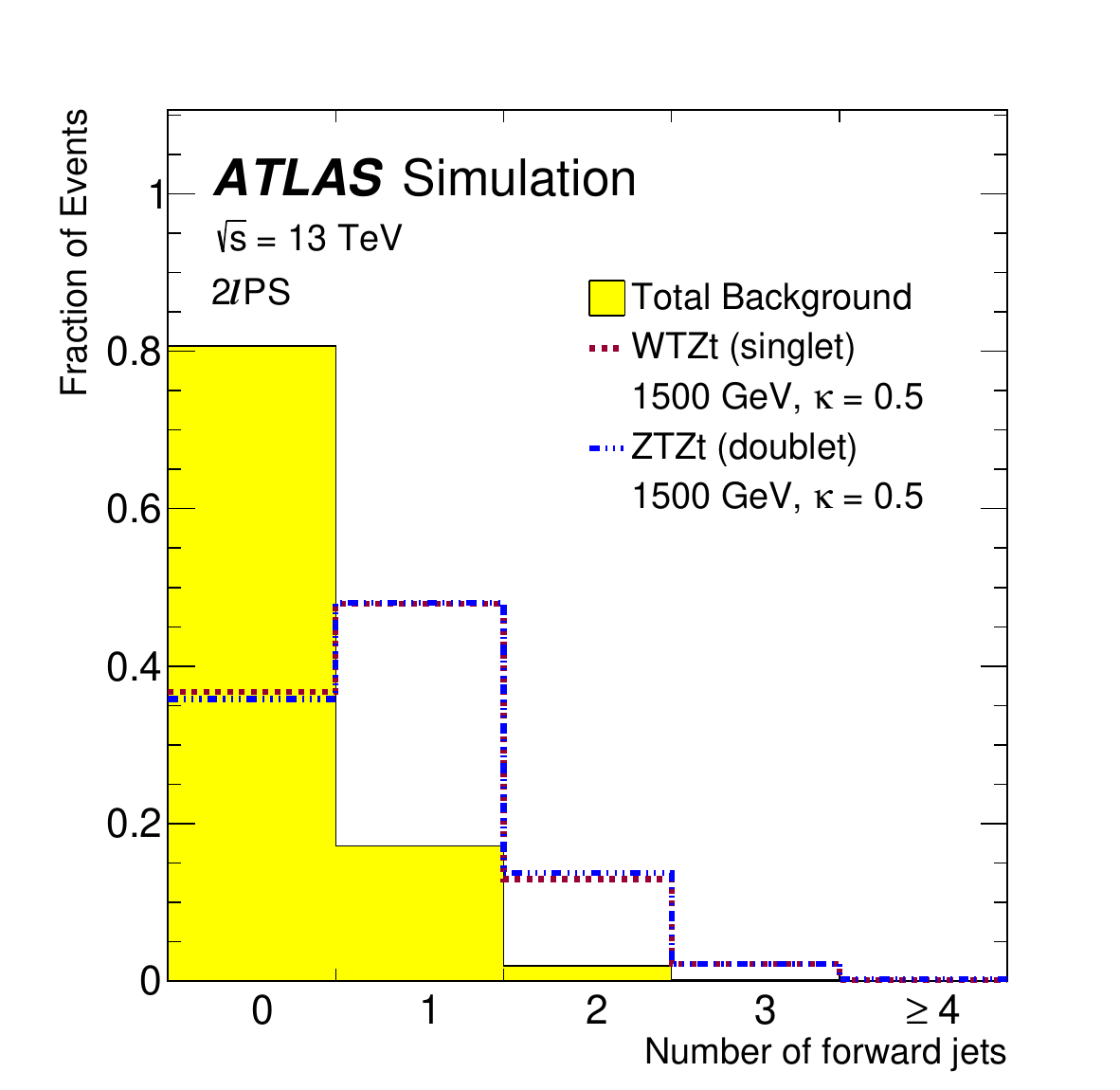}
\label{fig:2l_n_fwdjet_ps}
}
\subfloat[]{
\includegraphics[width=0.32\textwidth]{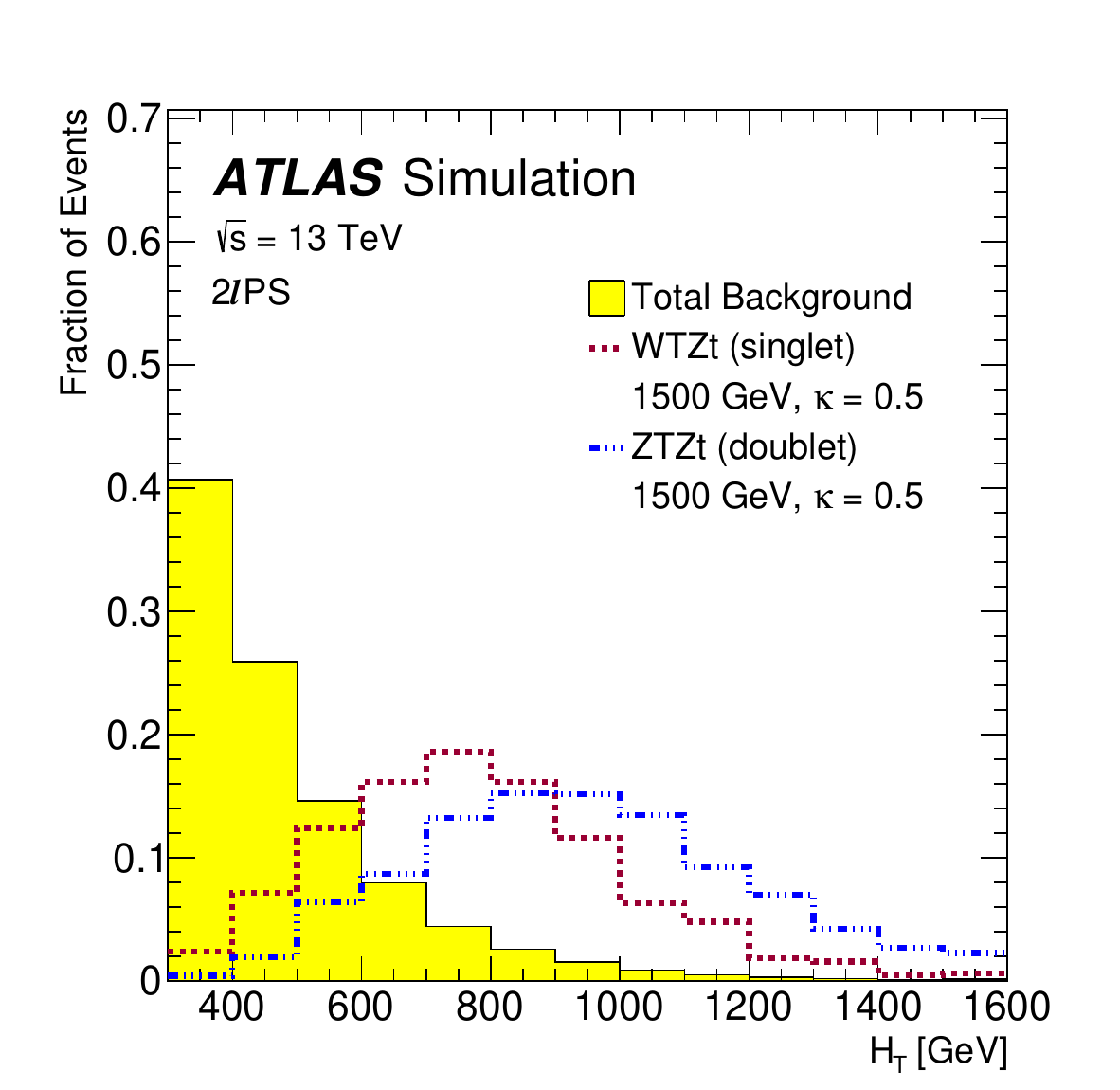}
\label{fig:2l_ht_jet_ps}
}
\caption[]{Simulated distributions of \subref{fig:2l_n_bjet_ps} $b$-tagged jet multiplicity, \subref{fig:2l_n_fwdjet_ps} forward jet multiplicity, and \subref{fig:2l_ht_jet_ps} \HT for total
background (solid area) and benchmark signal processes (dashed lines) in the \dilepreg{PS} region before reweighting is performed in jet multiplicity and $\HT + \met$ distributions. Each distribution is
separately normalized to unit area. The signal contributions are from $WTZt$ and $ZTZt$ processes with $M_T = 1.5$~\TeV and $\kappa =  0.5$ in the singlet and doublet representations respectively. The last bin in each distribution contains the overflow.
}
\label{fig:2l_lineshapes}
\end{figure}

Three control regions (CRs) are defined to improve the background modeling and reduce its uncertainties. Two validation regions (VRs) are defined to validate the background prediction in the distribution of the final discriminant.
The event selection criteria for these regions are summarized in Table~\ref{tab:2lregions}.
Each of the control regions is obtained by inverting exactly two of the three jet cuts of the \dilepreg{SR} shown in Table~\ref{tab:2lregions}.
The regions \dilepreg{CR1}, \dilepreg{CR2}, and \dilepreg{CR3} are defined to  have the same selection as \dilepreg{SR} in multiplicities of forward jets, $b$-tagged jets, and top-tagged jets respectively.
The inversion of the requirement of the presence of top-tagged jets requires the presence of at least one top-vetoed tagged jet.
Both the validation regions are required to have at least one top-tagged jet to maintain their kinematic similarity with the signal region.
Identified as \dilepreg{VR1} and \dilepreg{VR2}, they are made orthogonal to the \dilepreg{SR} by requiring zero $b$-tagged jets and zero forward jets respectively.

\begin{table}[h!]
\centering
\caption{Summary of selections applied to define the control, validation, and signal regions for the \dilep channel.
}
\resizebox{0.7\textwidth}{!}{\begin{tabular}{|c|c|c|c|c|c|c|}
\hline
& \dilepreg{CR1} & \dilepreg{CR2} & \dilepreg{CR3} & \dilepreg{VR1} & \dilepreg{VR2} & \dilepreg{SR} \\
\hline
\hline
\multirow{4}{*}{Preselection} & \multicolumn{6}{c|}{= 2 OS-SF leptons with $| m(\ell\ell)-m_Z | < 10$~\GeV}\\
& \multicolumn{6}{c|}{$\ptz > 200$~\GeV, $\HT > 300$~\GeV }\\
& \multicolumn{6}{c|}{$\ge 1$ vRC jet}\\
& \multicolumn{6}{c|}{$H_\mathrm{T}+E_\mathrm{T}^{\mathrm{miss}} < m_{\ell\ell J}$}\\
\hline
\hline
forward jets & {$\geq 1$} & {0} & {0} & {$\geq 1$} & {0} & {$\geq 1$} \\
\hline
$b$-tagged jets & {0} & {$\geq 1$} & {0} & {0} & {$\geq 1$} & {$\geq 1$}\\
\hline
top-tagged jets & {--} & {--} & {$\geq 1$} & {$\geq 1$} & {$\geq 1$} & {$\geq 1$}\\
\hline
top-vetoed jets & {$\geq 1$} & {$\geq 1$} & {--} & {--} & {--} & {--}\\
\hline
\end{tabular} }
\label{tab:2lregions}
\end{table}


\subsection{Trilepton channel}
\label{subsec:SP3l}


%
As the name of the channel suggests, events with at least three leptons passing the preselection are selected. Although the selection is inclusive of events with four or more leptons, their overall contribution is quite small $(< 5\%)$. The lepton with highest \pt that does not comprise the $Z$-boson candidate is labeled as the third lepton $(\ell_3)$ in this channel.
The main sources of background  are diboson processes and $t\bar t$+X, the latter being primarily composed of \ttbarZ and other small contributions from \ttbarW and $t\bar{t}t\bar{t}$ processes. Minor contributions result from background processes like single-top and \zjets.
To improve the quality of MC background modeling, a data-driven reweighting factor is determined for the \diboson and \ttbarV samples in this channel.
This reweighting is performed for the central jet multiplicity distribution for events passing the \trilepreg{PS} criteria.
These events are classified into separate regions according to the presence or absence of $b$-tagged jets in the event. A pair of simultaneous linear equations can be formulated for these two sets of data and MC distributions:
\begin{align}
n_{0,i}^{\mathrm{data}} &= \alpha_i^{VV} n_{0,i}^{VV} + \alpha_i^{ttX}n_{0,i}^{ttX} + n_{0,i}^{\mathrm{others}} \nonumber \\
n_{1,i}^{\mathrm{data}} &= \alpha_i^{VV} n_{1,i}^{VV} + \alpha_i^{ttX}n_{1,i}^{ttX} + n_{1,i}^{\mathrm{others}}
\label{eqn:3l-reweighting}
\end{align}
where $n^{\mathrm{data}}$ denotes the data yield and  $n^{VV}, n^{ttX},$ and $n^{\mathrm{others}}$ denote the predicted yields from $VV$, $t\bar{t}+X$, and other backgrounds respectively.
The subscripts $0/1$ refer to events with $0/\geq 1$ $b$-tagged jets and the index $i$ runs over the bins of the central  jet multiplicity distribution.
The $(\alpha_i^{VV}, \alpha_i^{ttX})$ factors for each bin are obtained by simultaneously solving the pair of
equations in Eq.~(\ref{eqn:3l-reweighting}) and applied to estimate the MC contributions of \diboson and \ttbarV processes.
The central jet multiplicity
distribution in the \trilepreg{PS} is shown in Figure~\ref{fig:3l_rwt}.
In the regions where the reweighting factors are derived, the overall contribution of the minor backgrounds is up to 29\%, reaching a bin-by-bin maximum of 37\%. The predicted signal contributions in these regions are small. For a benchmark $WTZt$ process with $M_T = 1.5$~\TeV and $\kappa = 0.5$ in the singlet representation, the bin-by-bin $S/B$ is less than 2\%.

\begin{figure}
\centering
\includegraphics[width=0.48\textwidth]{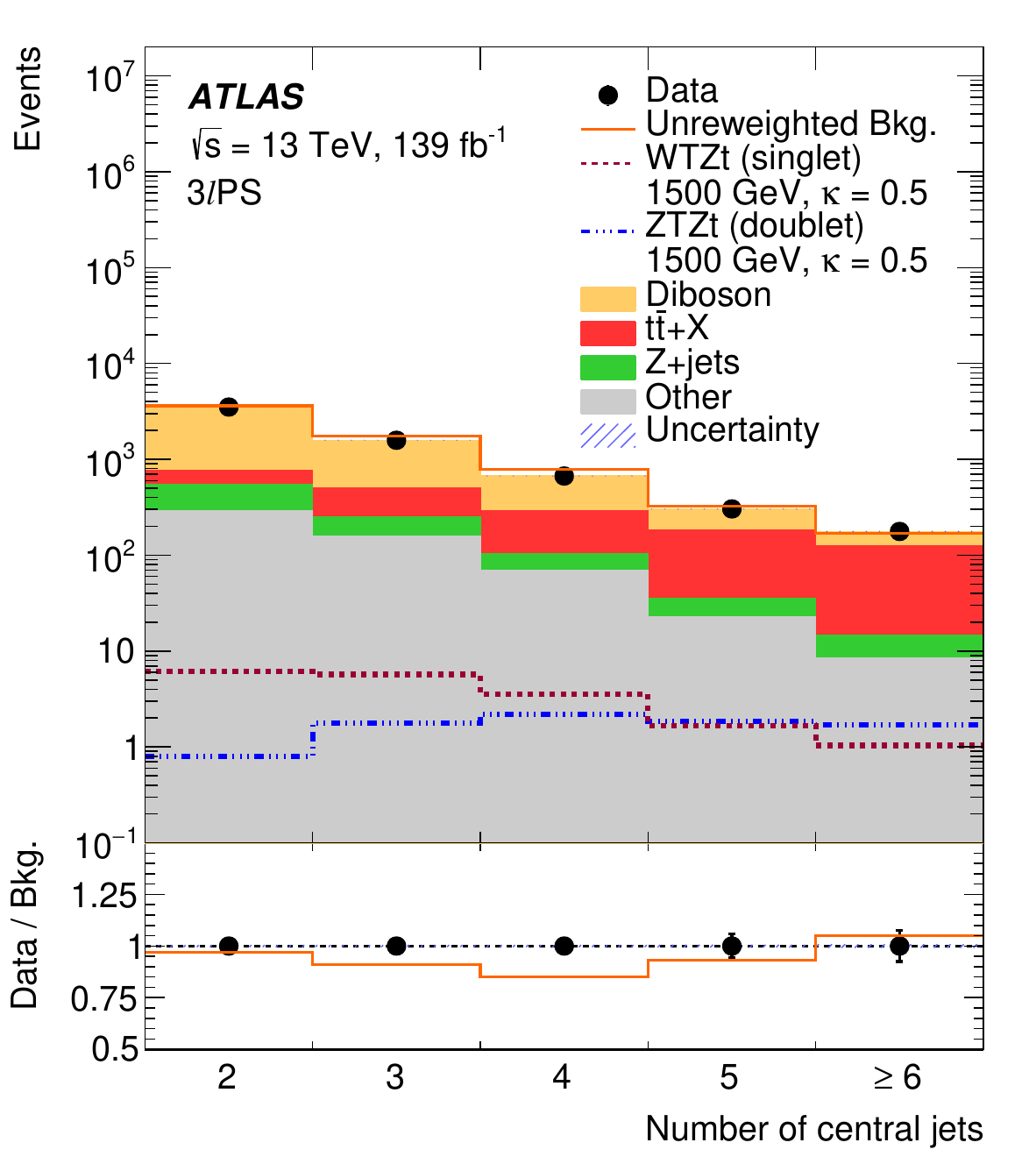}
\label{fig:3l_ncntjet_ps}
\caption[]{Distribution of central jet multiplicity  in the \trilepreg{PS} region after reweighting. The overlaid solid line shows the total background prediction before the reweighting and the corresponding data-over-background ratio in the bottom panel. The red and blue dotted lines show the signal contributions from $WTZt$ and $ZTZt$ processes with $M_T = 1.5$~\TeV and $\kappa =  0.5$ in singlet and doublet representations respectively.
}
\label{fig:3l_rwt}
\end{figure}

Figure~\ref{fig:3l_lineshapes} shows distributions of some of the kinematic variables for the expected background and benchmark signal processes in the \trilepreg{PS} region. Exploiting the distinctive signature of signal events, which are expected to have $b$-tagged and forward jets, the signal region in this channel (\trilepreg{SR}) requires events to have at least one of each of these jet types.
To further increase the signal-to-background ratio, additional requirements are applied on the $Z$-boson candidate \pt, $\ptz> 300$~\GeV and the leading lepton \pt, $\max(\pt(\ell)) > 200$~\GeV.
The decay products of $T$, the $Z$ boson and the top-quark, are expected to have large angular separation between them.
Requiring the azimuthal separation of the $Z$ candidate from the third lepton $(\Delta\phi(Z, \ell_3))$ to be at least $\frac{\pi}{2}$ significantly improves the signal purity in the signal region  (Figure~\ref{fig:3l_dphi_Zl_ps}).
For similar reasons, the azimuthal separation between the $Z$ candidate and the leading $b$-tagged jet, defined as the  $b$-tagged jet with the largest \pt which is expected to emerge from the decay of the top quark, is required to satisfy $\Delta\phi(Z, b_{\mathrm{lead}}) > \frac{\pi}{2}$.

\begin{figure}
\centering
\subfloat[]{
\includegraphics[width=0.32\textwidth]{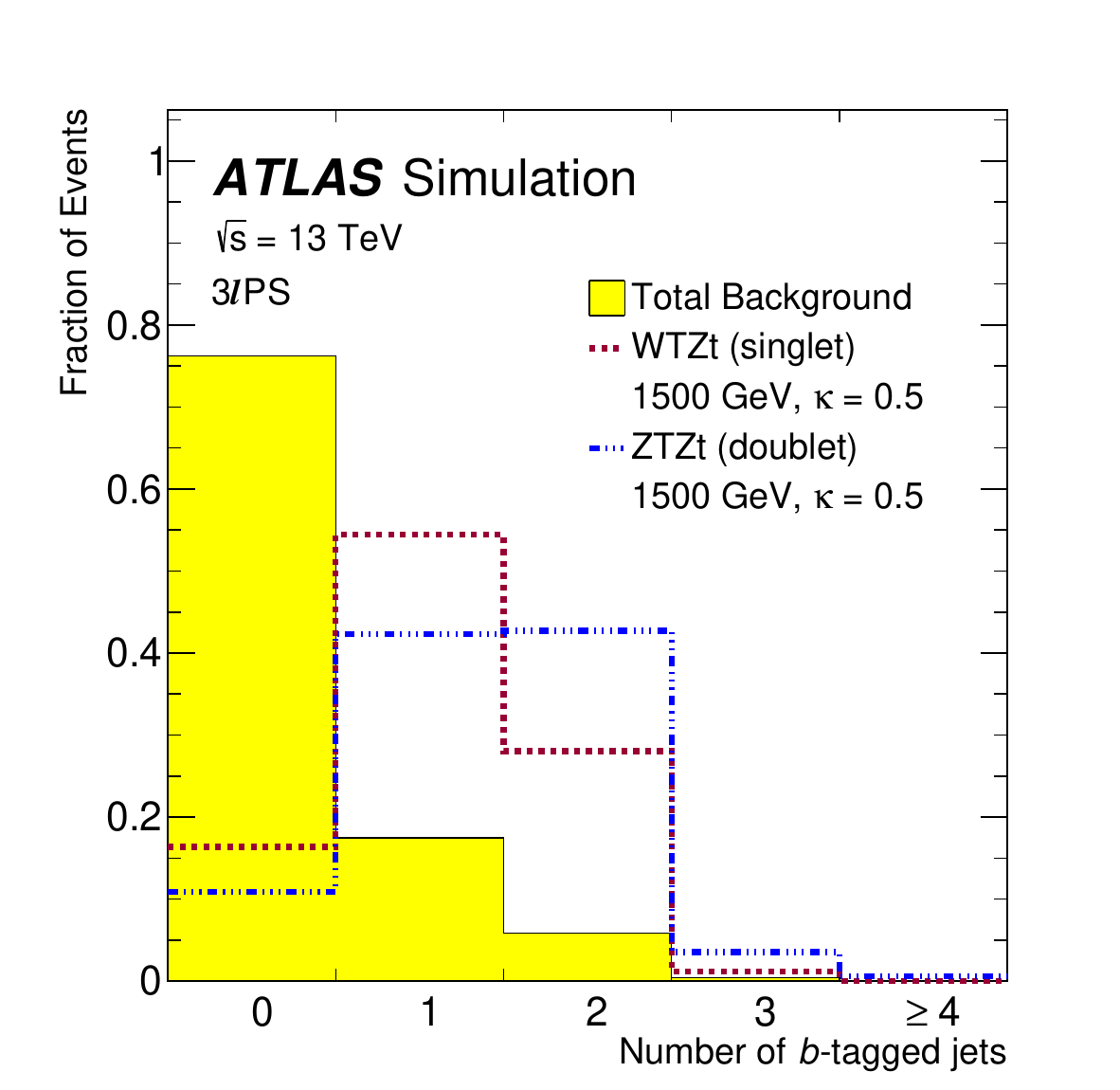}
\label{fig:3l_n_bjet_ps}
}
\subfloat[]{
\includegraphics[width=0.32\textwidth]{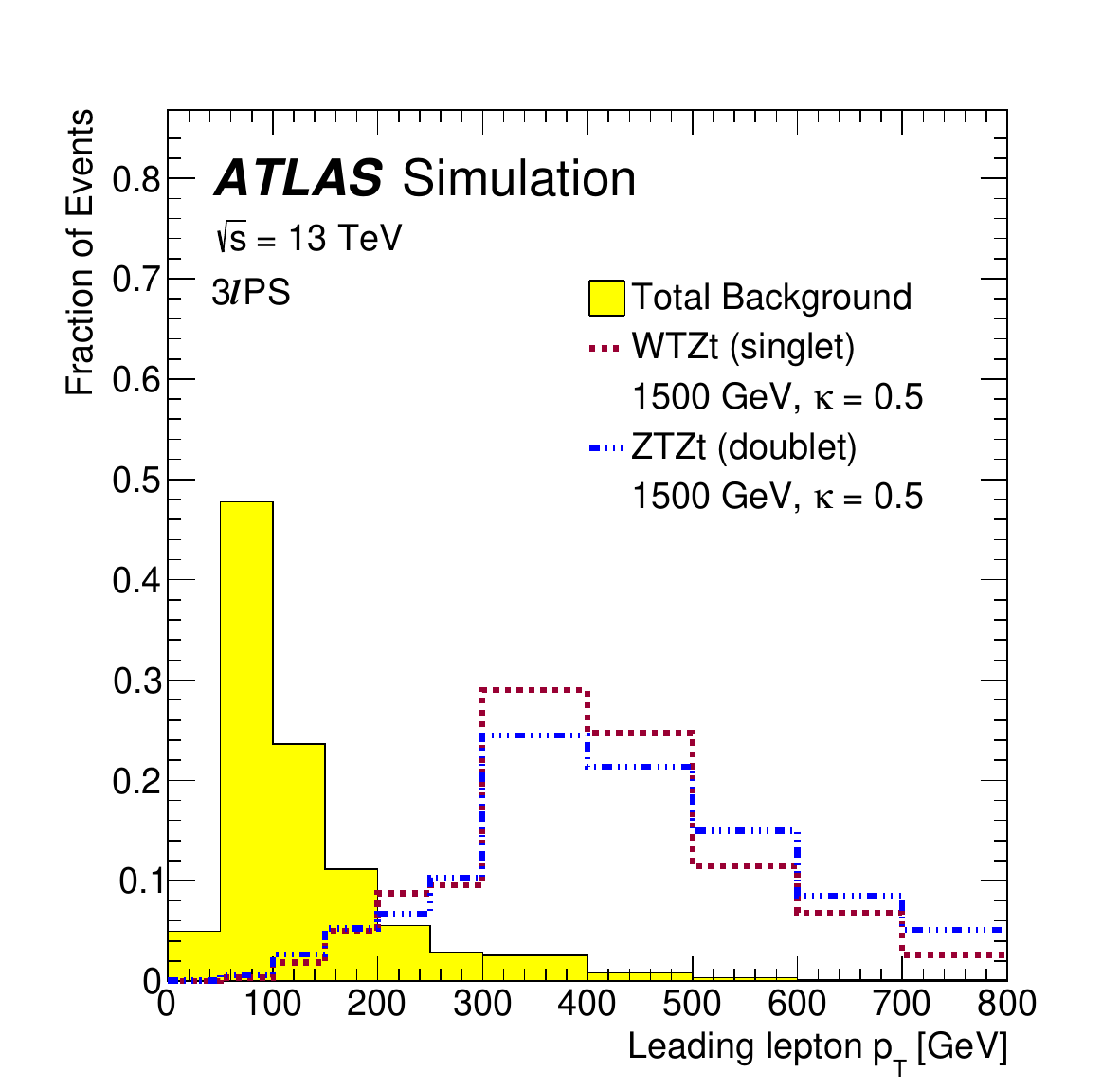}
\label{fig:3l_pt_lept_lead_ps}
}
\subfloat[]{
\includegraphics[width=0.32\textwidth]{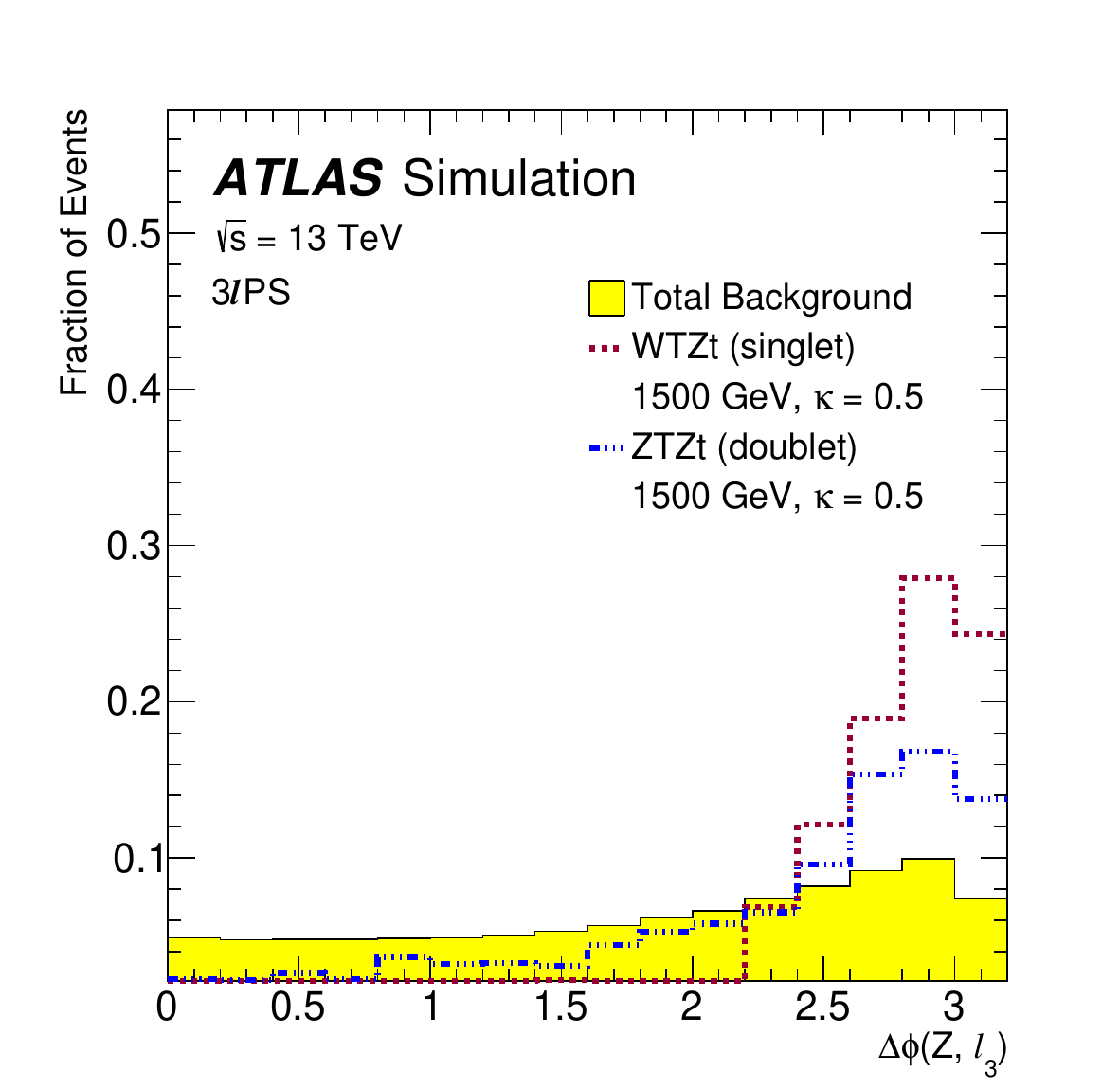}
\label{fig:3l_dphi_Zl_ps}
}
\caption[]{Simulated distributions of \subref{fig:3l_n_bjet_ps} $b$-tagged jet multiplicity, \subref{fig:3l_pt_lept_lead_ps} $p_{T}$ of the leading lepton,
and \subref{fig:3l_dphi_Zl_ps} $\Delta\phi$ between the $Z$ candidate and the third lepton for total background (solid area) and benchmark signal processes (dashed lines) in
the \trilepreg{PS} region before reweighting is performed in the central jet multiplicity distribution. Each distribution is separately normalized to unit area.
The signal contributions are from $WTZt$ and $ZTZt$ processes with $M_T = 1.5$~\TeV and $\kappa =  0.5$ in the singlet and doublet representations respectively.
The last bin in each distribution contains the overflow.}
\label{fig:3l_lineshapes}
\end{figure}

Finally, to allow for a consistent phenomenological interpretation,  the product of $H_{\mathrm{T}}$ and jet multiplicity, $H_\mathrm{T} \cdot n(\textrm{jets})$, is required to be less than 6~\TeV in the signal region to reduce potential contamination from pair production of VLQs~\cite{EXOT-2016-35}.

The statistical analysis in this channel also uses \ptz as the final discriminant. Three control regions and one validation region are defined in this channel. The diboson control region (\trilepreg{VV}) is defined by rejecting events with any $b$-tagged jet.
The mixed control region (\trilepreg{Mixed}) and the $t\bar{t}+X$ control region (\trilepreg{ttX}), are defined by accepting events with exactly one $b$-tagged jet and at least two $b$-tagged jets respectively. To maintain orthogonality with the signal region, both regions reject events with forward jets and require $\Delta\phi(Z, \ell_3) < 2.6$.
Finally, a validation region (\trilepreg{VR}) is defined to validate the background modeling  by  requiring the same set of selections on the $b$-tagged jet and forward jet multiplicities as the signal region, but made orthogonal to it by requiring that at least one of the $\Delta\phi$ requirements is reversed, i.e., $\Delta\phi(Z, \ell_3) < \frac{\pi}{2}$ or $\Delta\phi(Z, b_{\mathrm{lead}}) < \frac{\pi}{2}$. The definitions of these regions are summarized in Table  \ref{tab:3lregions}.

\begin{table}[h!]
\centering
\caption{Summary of selections applied to define the control, validation, and signal regions for the \trilep channel.}
\resizebox{0.9\textwidth}{!}{
\begin{tabular}{|c|c|c|c|c|c|}
\hline
& \trilepreg{VV} & \trilepreg{Mixed} & \trilepreg{ttX} & \trilepreg{VR} & \trilepreg{SR}\\
\hline\hline
\multirow{2}{*}{Preselection} & \multicolumn{5}{c|}{$\geq$ 3 leptons} \\
& \multicolumn{5}{c|}{$\geq$ 1 pair of OS-SF leptons with $|m(\ell\ell) - m_Z| < 10$ \GeV} \\
& \multicolumn{5}{c|}{$\geq 2$ central jets} \\
\hline\hline
$b$-tagged jets &       0 &     1 &     $\geq 2$ &      {$\geq 1$} &    {$\geq 1$}\\
\hline
forward jets    &       -- &     0 &            0 &      {$\geq 1$} &    {$\geq 1$}\\
\hline
\multirow{2}{*}{$\Delta\phi$ selections}      &       \multirow{2}{*}{--} & \multirow{2}{*}{$\Delta\phi(Z, \ell_3) < 2.6$} & \multirow{2}{*}{$\Delta\phi(Z, \ell_3) < 2.6$}
& $\Delta\phi(Z, \ell_3) < \frac{\pi}{2}$ \textbf{OR}
& $\Delta\phi(Z, \ell_3) > \frac{\pi}{2}$ \textbf{AND}\\
&   &   &   & $\Delta\phi(Z, b_{\mathrm{lead}}) < \frac{\pi}{2}$   & $\Delta\phi(Z, b_{\mathrm{lead}}) > \frac{\pi}{2}$\\
\hline
\multirow{3}{*}{other selections} & \multirow{3}{*}{--}    & \multirow{3}{*}{--}   & \multirow{3}{*}{--}   & \multirow{3}{*}{--}  & $\max(p_{\mathrm{T}}(\ell)) > 200$~\GeV\\
&                           &   &   &   & $p_{\mathrm{T}}(\ell\ell) > 300$~\GeV\\
&                           &   &   &   & $H_\mathrm{T} \cdot n(\textrm{jets})<$  6 \TeV\\

\hline
\end{tabular}
}
\label{tab:3lregions}
\end{table}



%

%
%
%

%
\section{Systematic uncertainties}
\label{sec:syst}

%
%
%
%

Uncertainties from  experimental and theoretical sources are introduced in the normalization or shape of the final discriminant in the two analysis channels.
Experimental uncertainties include effects on the electron energy scale and energy resolution,
the muon momentum scale and resolution, and uncertainties in the data-to-MC correction
factors for the electron and muon trigger, reconstruction, identification, and isolation
efficiencies~\cite{EGAM-2018-01,MUON-2018-03}.
Jet energy scale and resolution uncertainties are also included,
as obtained from studies in data and simulation~\cite{jes}.
Flavor-tagging uncertainties include uncertainties in the \bjet tagging, \cjet mis-tagging, and light-jet mis-tagging efficiencies, and uncertainties due to extrapolations to regions not covered by the data used for the efficiency measurements~\cite{FTAG-2018-01,FTAG-2020-08,FTAG-2019-02}.
Subdominant uncertainties include uncertainties related to the soft term in
the \met\ calculation~\cite{PERF-2016-07} and to the \met energy scale and resolution,
uncertainties in the reweighting of the MC event samples to match the pileup conditions in data and a 1.7\%~\cite{ATLAS-CONF-2019-021}
uncertainty in the integrated luminosity of the 2015--2018 data sample, obtained using the LUCID-2 detector \cite{LUCID2} for the primary luminosity measurements.

Theoretical uncertainties include cross-section and other modeling uncertainties for all
background samples. The cross-section uncertainties considered are between 5 -- 6\% for the
\zjets, \ttbar, and \VV samples~\cite{STDM-2016-01,Czakon:2011xx,campbell1999update} and between 10 -- 12\%  for the \ttbarZ sample~\cite{TOPQ-2018-08}.
Uncertainties associated with missing higher orders are estimated by varying
the renormalization and factorization scales by factors of 0.5 and 2 and then,
constructing upward and downward varying envelopes by considering the largest positive and negative fluctuation on a bin-by-bin basis.
PDF uncertainties when comparing distributions obtained with events generated with  different PDF sets are obtained using the PDF4LHC recommendations~\cite{Butterworth:2015oua}.
For the \dilep channel, the \zjets scale and PDF  variations are decorrelated between Z+LF and Z+HF events. %
For the \trilep channel, the \diboson scale-variation uncertainties are broken down into their normalization and shape components, and treated as uncorrelated across regions with different $b$-jet multiplicities.

Uncertainties due to the choice of generator
or showering algorithm are estimated by comparing additional
samples from alternative generators.
Generator uncertainties for \ttbar+$X$ are obtained from comparing the nominal \MGNLO samples with the alternative \SHERPA samples. For the \trilep channel where \zjets is a minor background and mainly contributes via jets mis-identified as leptons, generator uncertainties are estimated by comparing nominal samples with alternative \MGNLO samples and only taking the shape difference into account. For the \dilep channel where \zjets is the main background, generator uncertainties are taken into account by comparing the \SHERPA[2.2.1] sample with the \SHERPA[2.2.11] sample where an independent reweighting based on jet multiplicity and $\HT + \MET$ distribution, as explained in Section~\ref{subsec:SP2l}, is  applied on the latter.
To estimate the uncertainties due to the modeling of initial-state radiation, final-state radiation, and the choice of parton shower algorithm, either additional event weights for the nominal samples or alternative MC samples  were produced. These uncertainties were calculated by comparing the nominal MC distribution with the distributions obtained from these additional event weights or the alternative samples.

A conservative uncertainty of 30\% on the fraction of \VV and \zjets events with heavy-flavor jets is applied in both the channels. This uncertainty is based on variations of the factorization and renormalization scales and
the multijet merging scale~\cite{HIGG-2020-23}. Also, a shape uncertainty is introduced in the \trilep channel for \diboson and \ttbar+X by comparing the distributions before and after implementing jet multiplicity based reweighting explained in Section~\ref{subsec:SP3l}.
In the \dilep channel,
bin-by-bin variations in the reweighting factors due to the statistical fluctuations in the MC samples are calculated and these variations are used to obtain upward and downward fluctuating envelopes for the fit distributions.
Shape uncertainties extracted from these fluctuations are applied on the \zjets samples for the light and heavy-flavor components.

Events with jets misidentified as leptons can potentially arise in the \dilep channel mainly from the \ttbar process and in the \trilep channel from the \ttbar or $Z$+jets processes.
After the $m(\ell\ell)$ and $\pt(\ell\ell)$ cuts, these contributions are estimated to be less than 2\% in the signal regions. Following the prescription in Ref.~\cite{EXOT-2016-35}, a conservative 25\% uncertainty is applied on the normalization of the  \ttbar (\ttbar and $Z$+jets) events in the \dilep (\trilep) channel.
This uncertainty is decorrelated across the control and signal regions in the \dilep and \trilep channels.


%

%
\section{Statistical analysis and results}
\label{sec:stat-result}

%
%
A binned profile likelihood fit in the discriminating variable $\pt(\ell\ell)$ is performed in both channels to test if the data is compatible with the background-only hypothesis.
The uncertainties are introduced in the likelihood as nuisance parameters (NP), which are constrained using Gaussian priors.
Additional NPs are included
to take into account the statistical uncertainties in each bin for each event category due to the limited size of the simulated samples~\cite{barlow1993fitting}.
The likelihood function $L(\mu,\vec{\theta})$
is constructed as a product of Poisson probabilities for each bin of the discriminating variable in the control and signal regions,

\begin{equation}
L(\mu,\vec{\theta}) = \Pi_{i=1}^{N_{\mathrm{bins}}}\mathrm{Pois}\left(n_{i}^{\mathrm{data}}|b_i + \mu s_i\right) \times P(\vec{\theta}),
\label{eqn:likelihood}
\end{equation}
where $N_{\mathrm{bins}}$ is the total number of bins in control and signal regions.
The parameters, $n_{i}^{\mathrm{data}}, b_i, s_i$, represent the number of observed data, expected background and signal events respectively in the $i$-th bin.
The product of prior distributions for the NPs, $\left(\vec{\theta}\right)$, is given by $P(\vec{\theta})$.
The expected number of events in each bin depends on these NPs as well as the signal strength, $\mu$, which acts as a scaling factor for the signal cross-section.

The  background and signal templates are fitted to the observed data to obtain the best estimate of signal and
background contributions in the control and signal regions.
The fitted values of $\mu,\vec{\theta}$ are propagated to the validation regions to test the quality of the fit and the corresponding background modeling.
The background-only \textit{post-fit} distributions, obtained by maximizing the likelihood function in Eq.~(\ref{eqn:likelihood}) with a fixed choice of $\mu = 0$, are shown for the \dilep and \trilep control, validation, and signal regions in Figures~\ref{fig:2l_CRs}~and~\ref{fig:3l_CRs} respectively.
The expected pre-fit signal yields for the $WTZt$ and $ZTZt$ processes with $M_T = 1.5$~TeV and $\kappa = 0.5$ for the singlet and doublet representations of $T$ are also shown in these figures.
The contribution of different background processes, expected signal yields, and observed data in different regions for the \dilep and \trilep channels are summarized in Tables \ref{tab:dilep} and \ref{tab:trilep} respectively.
The post-fit background contribution shows reasonable agreement in all regions with the observed data within the uncertainties.

\begin{figure}
\centering
\subfloat[]{
\includegraphics[width=0.32\textwidth]{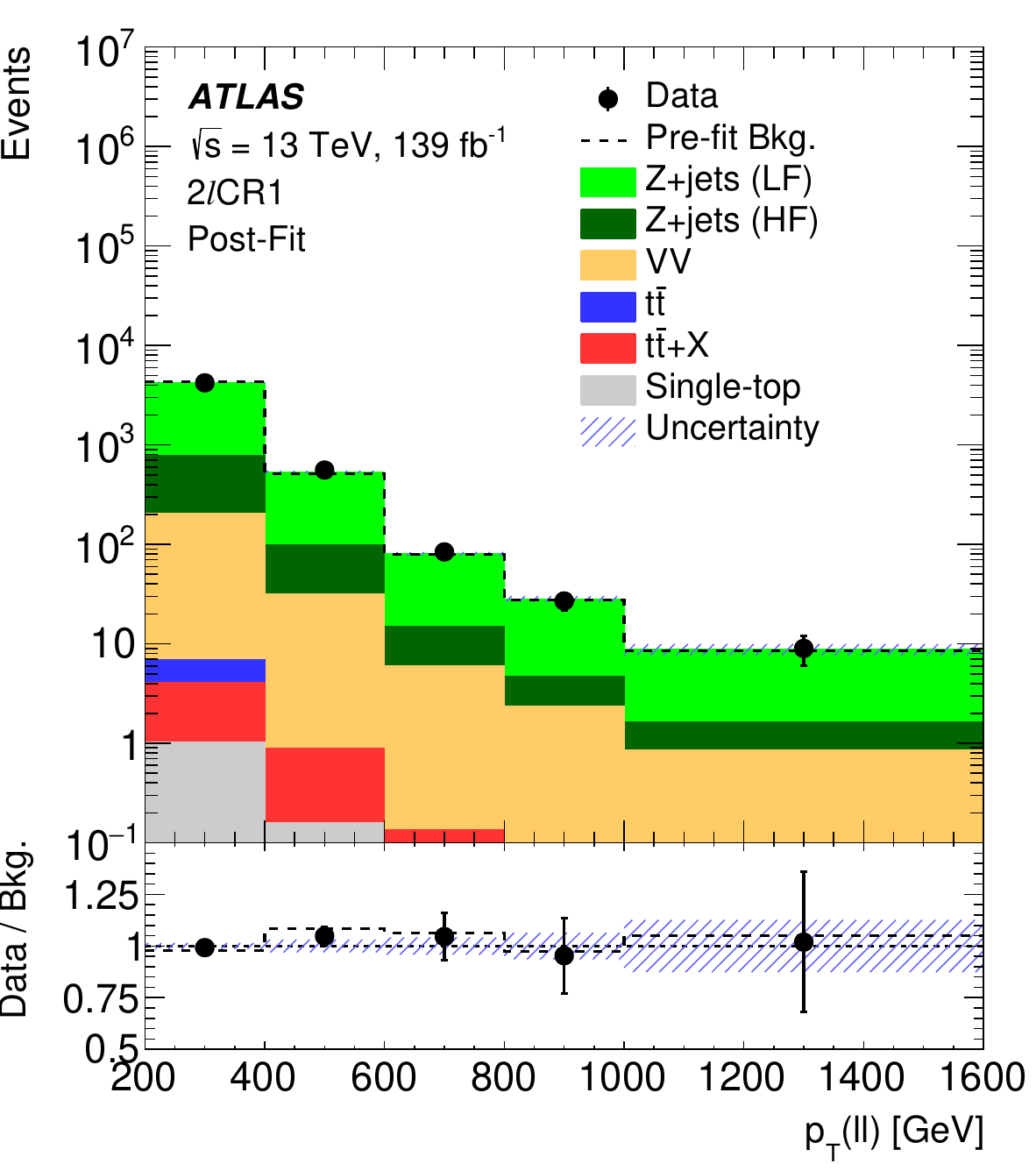}
\label{fig:2l_CR1}
}
\subfloat[]{
\includegraphics[width=0.32\textwidth]{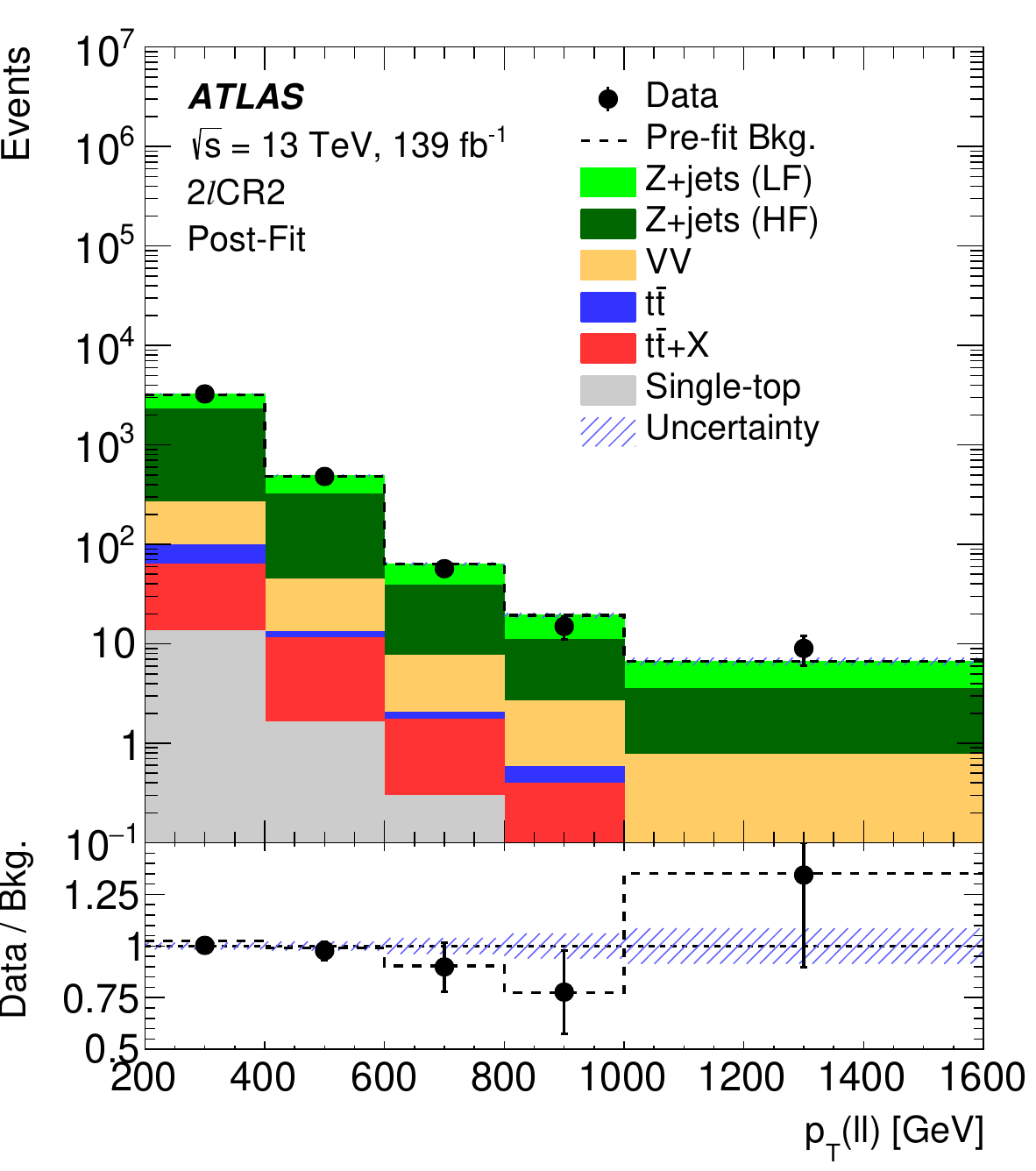}
\label{fig:2l_CR2}
}
\subfloat[]{
\includegraphics[width=0.32\textwidth]{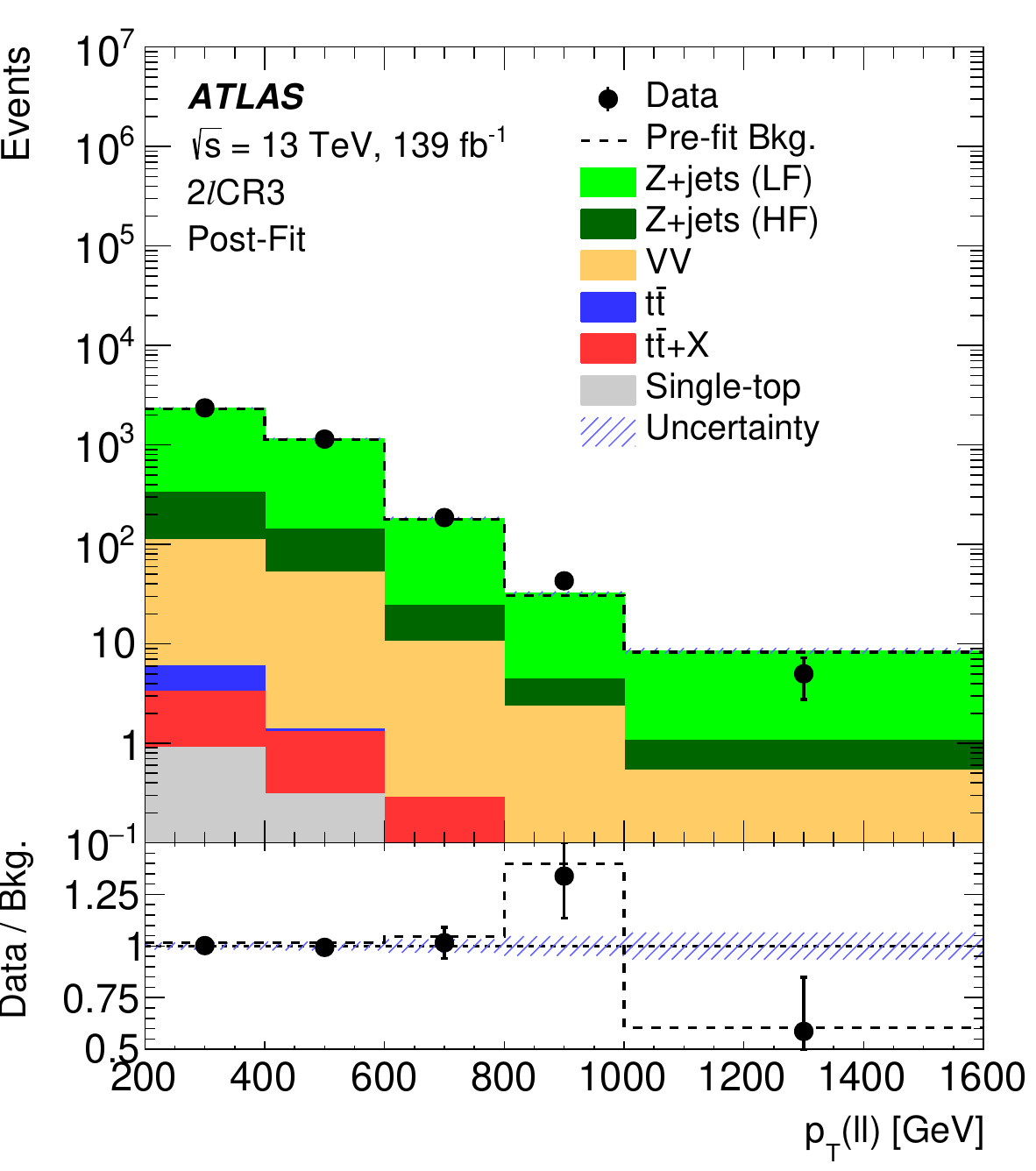}
\label{fig:2l_CR3}
} \\
\subfloat[]{
\includegraphics[width=0.32\textwidth]{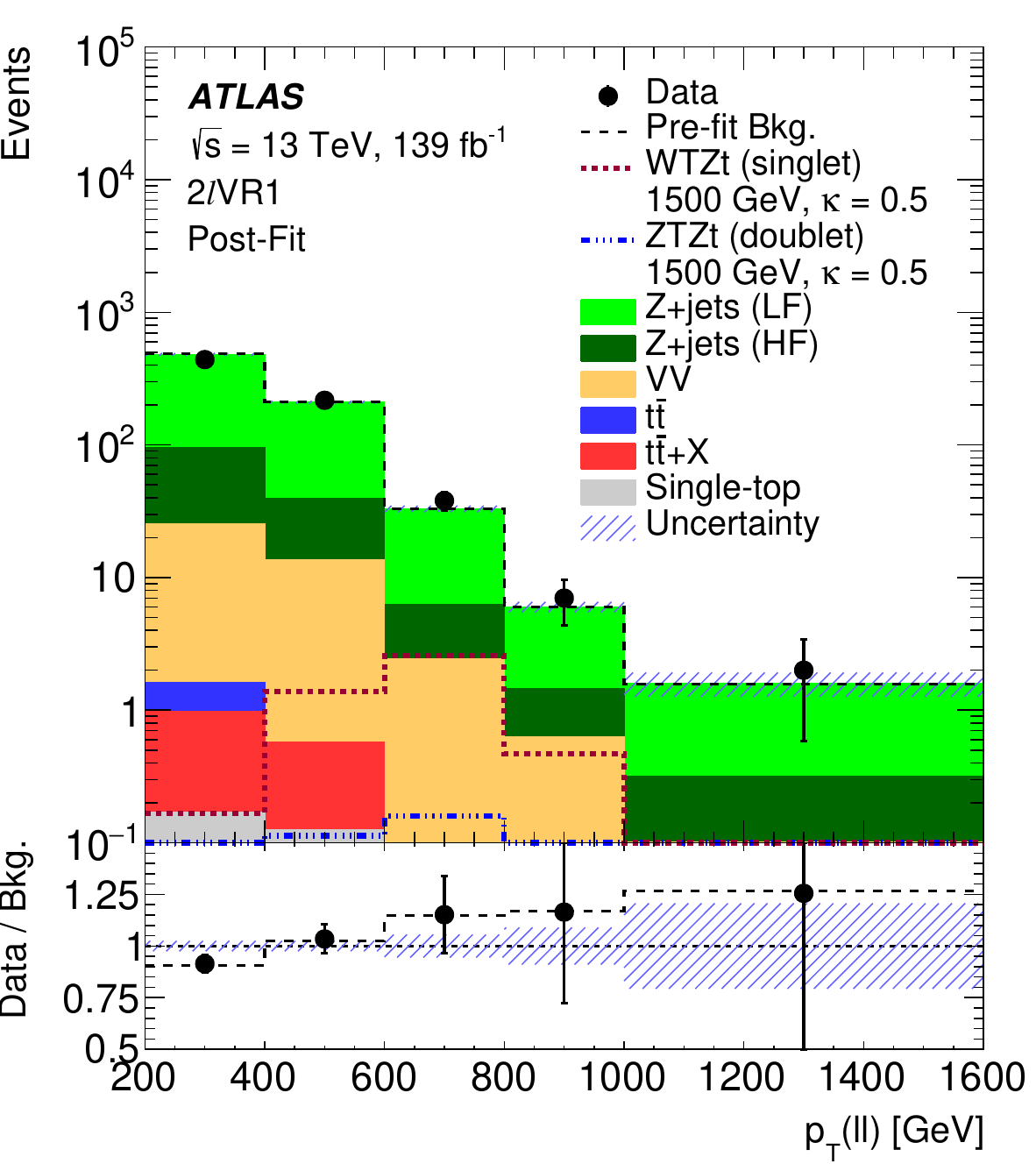}
\label{fig:2l_VR1}
}
\subfloat[]{
\includegraphics[width=0.32\textwidth]{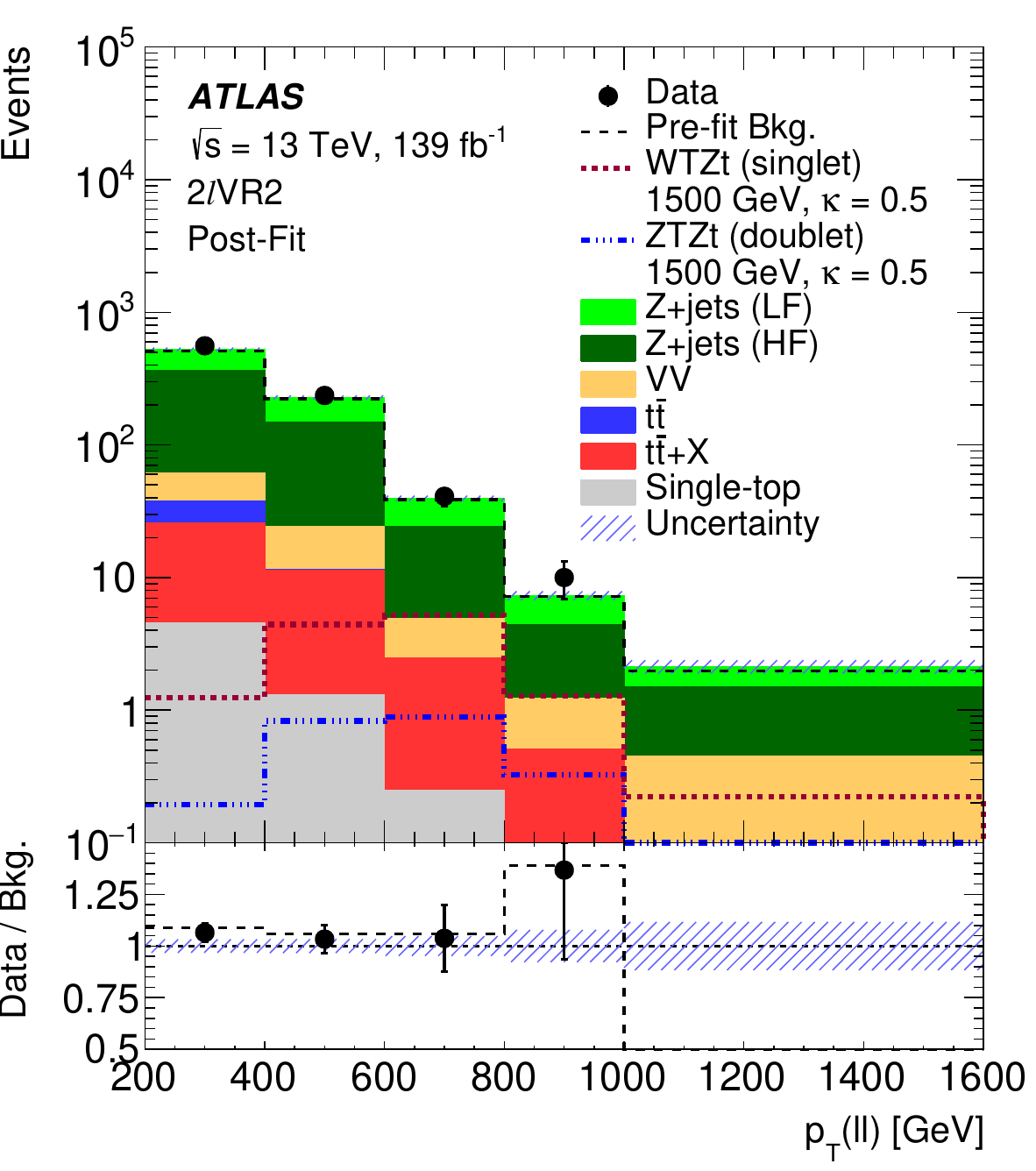}
\label{fig:2l_VR2}
}
\subfloat[]{
\includegraphics[width=0.32\textwidth]{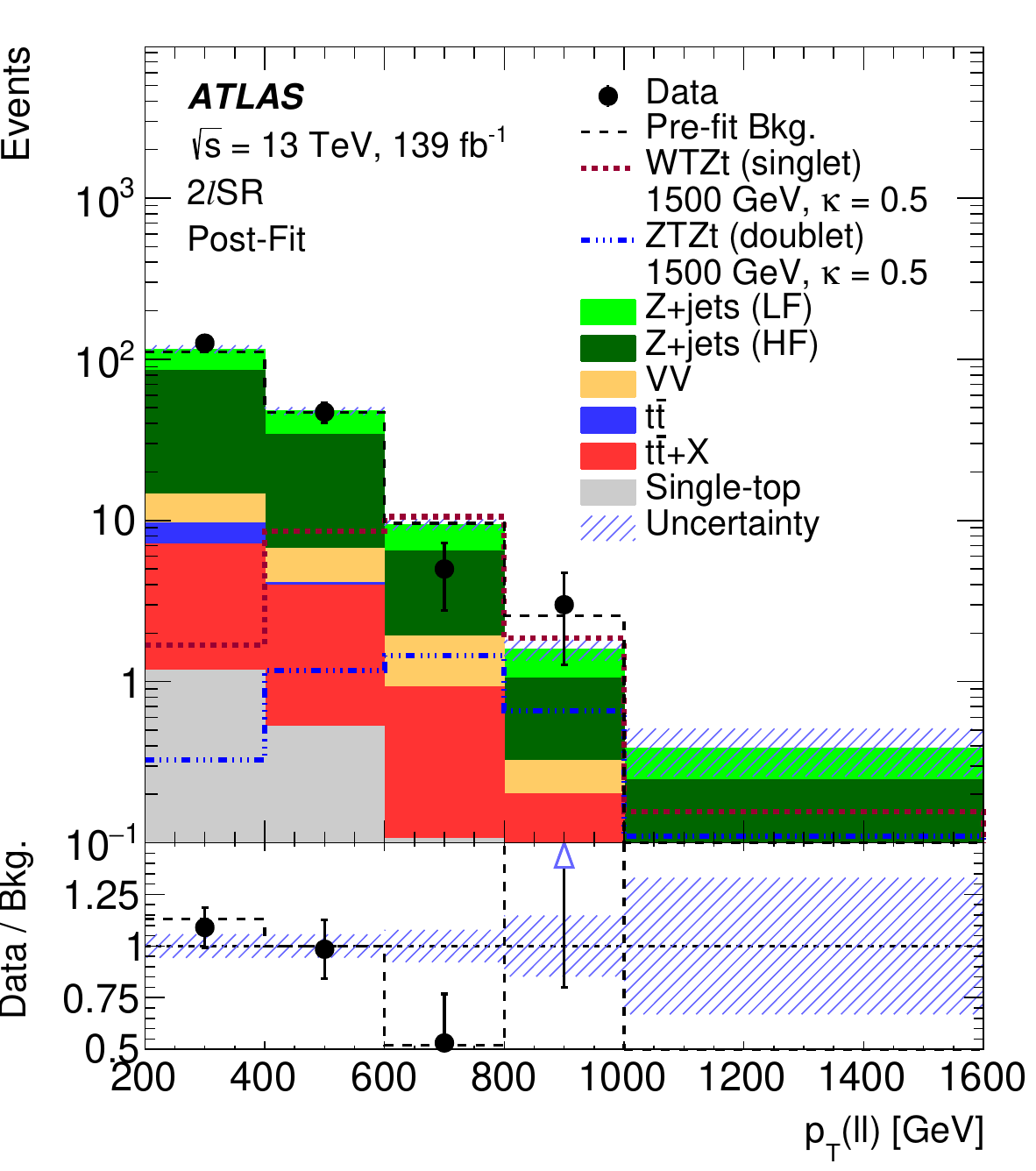}
\label{fig:2l_SR}
}
\caption[]{Distribution of the final discriminant, \ptz, for the \dilep channel in the control regions
\protect\subref{fig:2l_CR1} \dilepreg{CR1},
\protect\subref{fig:2l_CR2} \dilepreg{CR2},
\protect\subref{fig:2l_CR3} \dilepreg{CR3}, the validation regions
\protect\subref{fig:2l_VR1} \dilepreg{VR1},
\protect\subref{fig:2l_VR2} \dilepreg{VR2}, and the signal region
\protect\subref{fig:2l_SR} \dilepreg{SR}.
The distributions are shown after the background-only fit in the \dilep channel. The black, dashed line shows the total pre-fit background distribution and the corresponding data-over-background ratio in the bottom panel. The blue arrows in the bottom panel indicate points that are outside the vertical range of the figure. The overlaid red and blue dotted lines in  \protect\subref{fig:2l_VR1}-\protect\subref{fig:2l_SR} show the expected signal contributions in the corresponding regions from $WTZt$ and $ZTZt$ processes  with $M_T = 1.5$~\TeV and $\kappa = 0.5$ in singlet and doublet representations respectively. Expected signal contributions in the control regions are negligible. The last bin in each distribution contains the overflow.
}
\label{fig:2l_CRs}
\end{figure}


\begin{table}[ht!]
\centering
\caption{Observed number of events in data and post-fit number of background events in the control, validation, and signal regions for the \dilep channel. The fit was performed under a background-only hypothesis. Statistical uncertainties from the limited size of MC samples and systematic uncertainties are added in quadrature. Systematic uncertainties take the correlations among nuisance parameters into account. The $WTZt$ and $ZTZt$ signal yields correspond to pre-fit yields.
}
\resizebox{\textwidth}{!}{\begin{tabular}{|c|c|c|c|c|c|c|}

\hline

& \dilepreg{CR1}
& \dilepreg{CR2}
& \dilepreg{CR3}
& \dilepreg{VR1}
& \dilepreg{VR2}
& \dilepreg{SR}  \\
\hline
$WTZt$ (singlet)    & \multirow{2}{*}{3.9 $\pm$ 0.5}
& \multirow{2}{*}{9.8 $\pm$ 0.9}
& \multirow{2}{*}{3.6 $\pm$ 0.5}
& \multirow{2}{*}{4.7 $\pm$ 0.6}
& \multirow{2}{*}{12.4 $\pm$ 1.1}
& \multirow{2}{*}{22.8 $\pm$ 1.7} \\
$M_T = 1.5$~TeV, $\kappa=0.5$ & & & & & & \\
\hline
$ZTZt$ (doublet)    & \multirow{2}{*}{0.34 $\pm$ 0.04}
& \multirow{2}{*}{1.64 $\pm$ 0.16}
& \multirow{2}{*}{0.19 $\pm$ 0.04}
& \multirow{2}{*}{0.32 $\pm$ 0.05}
& \multirow{2}{*}{2.28 $\pm$ 0.22}
& \multirow{2}{*}{3.72 $\pm$ 0.24} \\
$M_T = 1.5$~TeV, $\kappa=0.5$ & & & & & & \\
\hline
\hline
\zjets (LF) & 3980 $\pm$ 90
& 1170 $\pm$ 110
& 3220 $\pm$ 80
& 589 $\pm$ 16
& 258 $\pm$ 27
& 48 $\pm$ 6 \\
\zjets (HF) & 660 $\pm$ 60
& 2330 $\pm$ 140
& 330 $\pm$ 50
& 101 $\pm$ 9
& 452 $\pm$ 28
& 103 $\pm$ 10 \\
\diboson    & 238 $\pm$ 35
& 206 $\pm$ 25
& 169 $\pm$ 26
& 40 $\pm$ 6
& 40 $\pm$ 7
& 8.7 $\pm$ 1.6 \\
\ttbar      & 2.8 $\pm$ 3.4
& 38 $\pm$ 20
& 2.8 $\pm$ 1.9
& 0.6 $\pm$ 0.6
& 12 $\pm$ 5
& 2.6 $\pm$ 1.6 \\
\ttbar+X    & 3.9 $\pm$ 1.0
& 61 $\pm$ 23
& 3.8 $\pm$ 0.8
& 1.34 $\pm$ 0.28
& 34 $\pm$ 12
& 10 $\pm$ 4 \\
Single-top  & 1.23 $\pm$ 0.17
& 15.6 $\pm$ 1.8
& 1.27 $\pm$ 0.15
& 0.33 $\pm$ 0.09
& 6.1 $\pm$ 0.8
& 1.85 $\pm$ 0.26 \\
\hline
Total background    & 4886 $\pm$ 70 %
& 3821 $\pm$ 60 %
& 3727 $\pm$ 60 %
& 732 $\pm$ 16
& 802 $\pm$ 24
& 174 $\pm$ 9 \\ %
\hline
Data    & 4887
& 3818
& 3735
& 704
& 846
& 181 \\
\hline
\end{tabular}}
\label{tab:dilep}
\end{table}


\begin{figure}
\centering
\subfloat[]{
\includegraphics[width=0.32\textwidth]{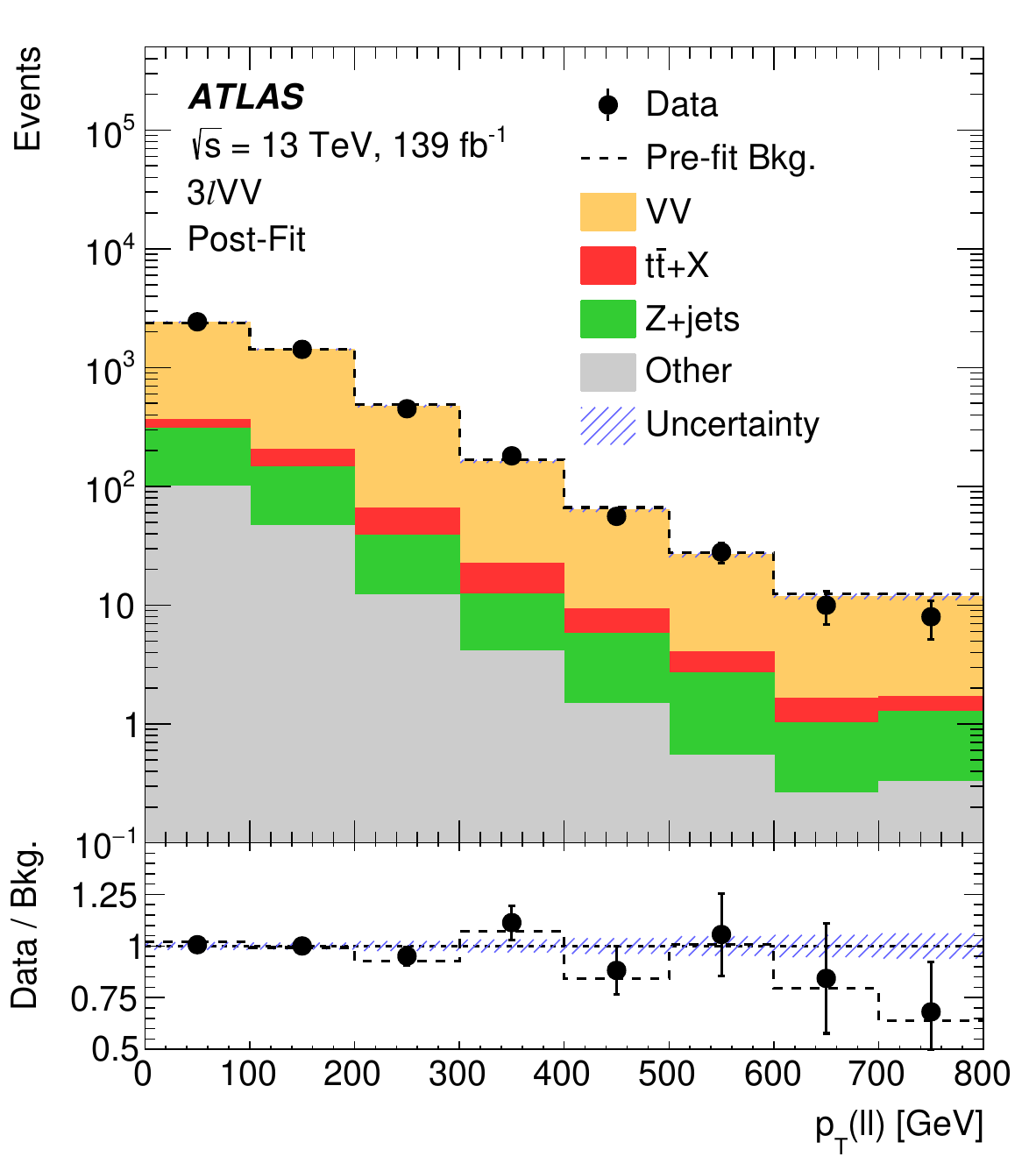}
\label{fig:3l_CR1}
}
\subfloat[]{
\includegraphics[width=0.32\textwidth]{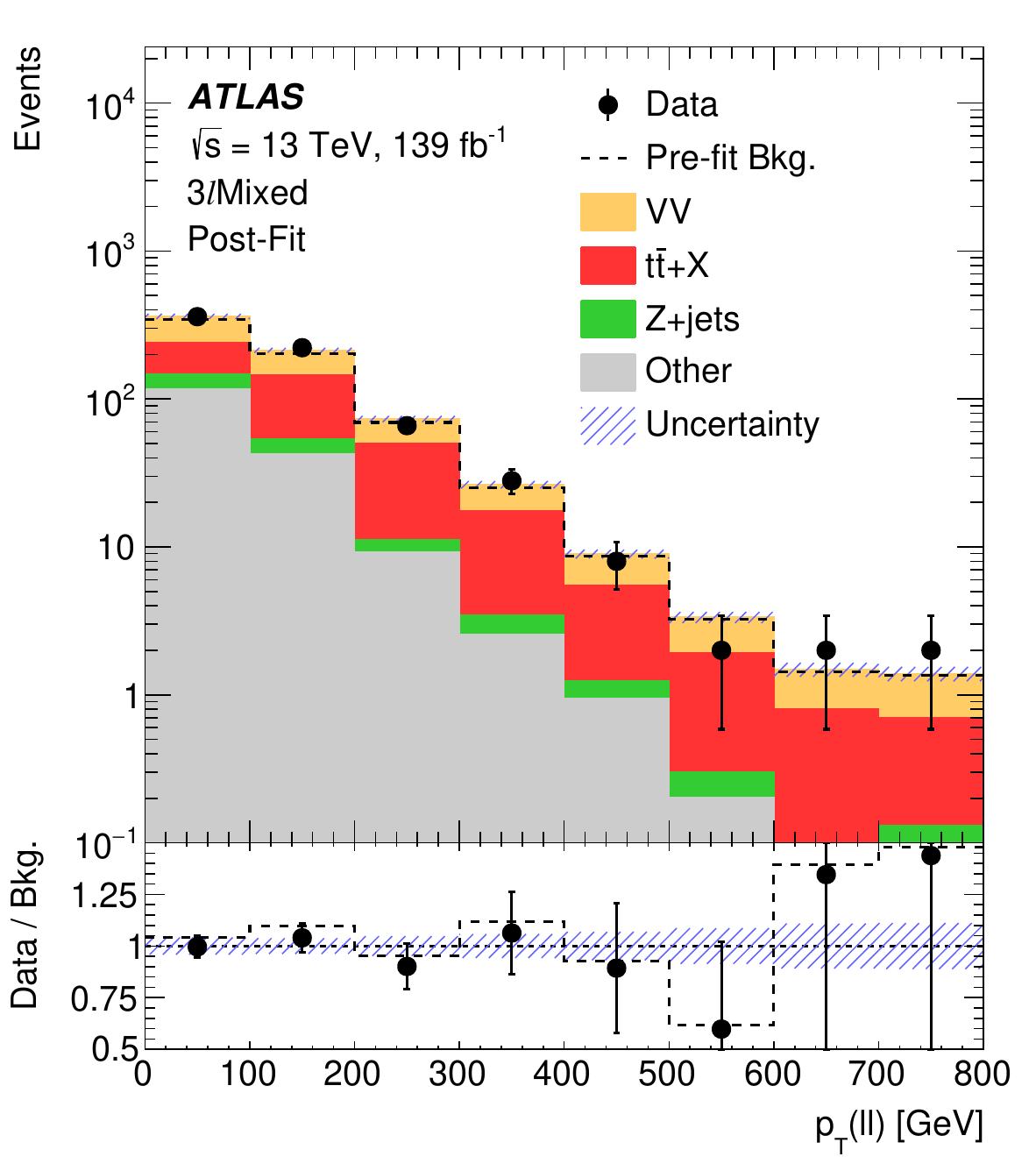}
\label{fig:3l_CR2}
}
\subfloat[]{
\includegraphics[width=0.32\textwidth]{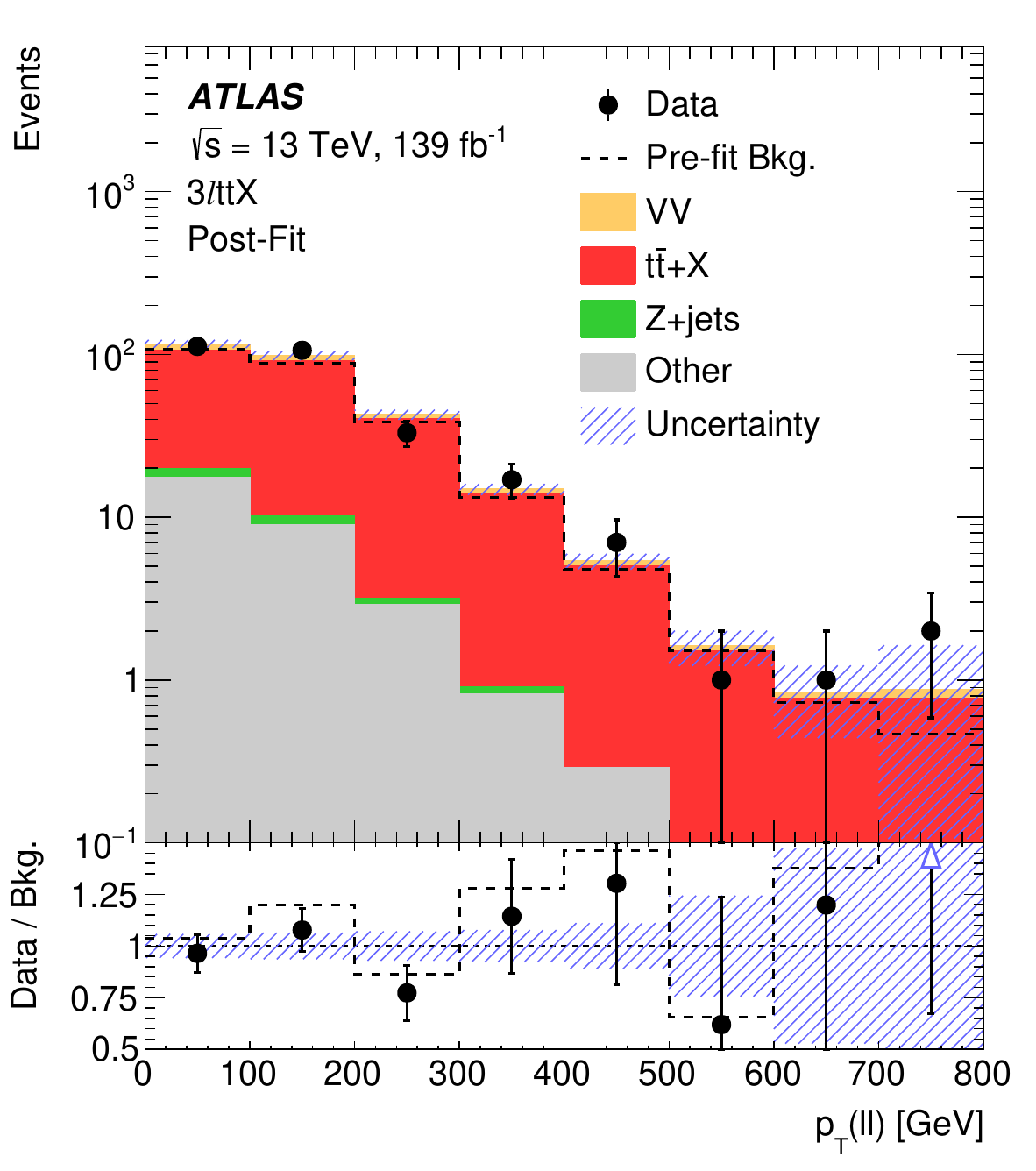}
\label{fig:3l_CR3}
} \\
\subfloat[]{
\includegraphics[width=0.32\textwidth]{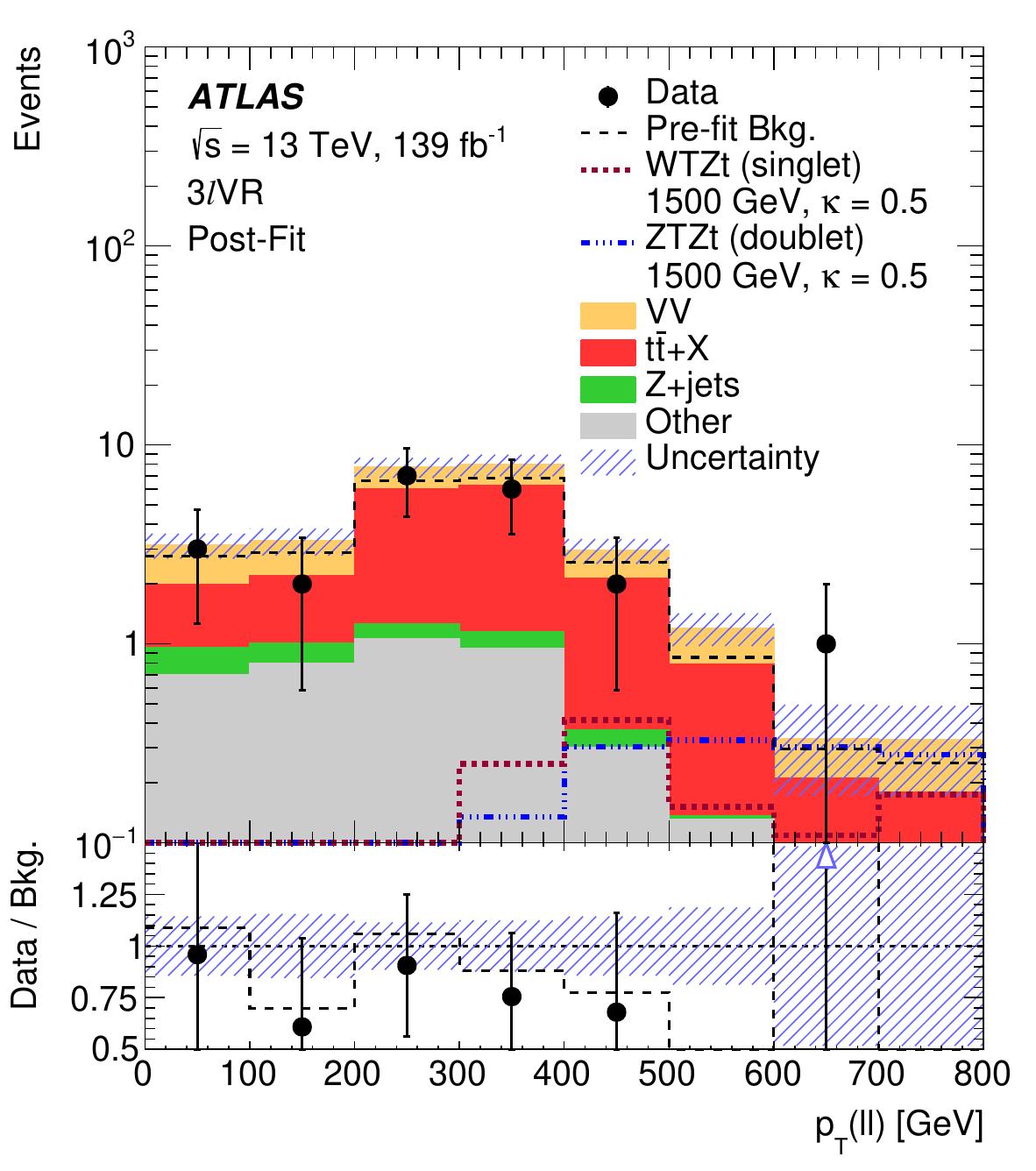}
\label{fig:3l_VR}
}
\subfloat[]{
\includegraphics[width=0.32\textwidth]{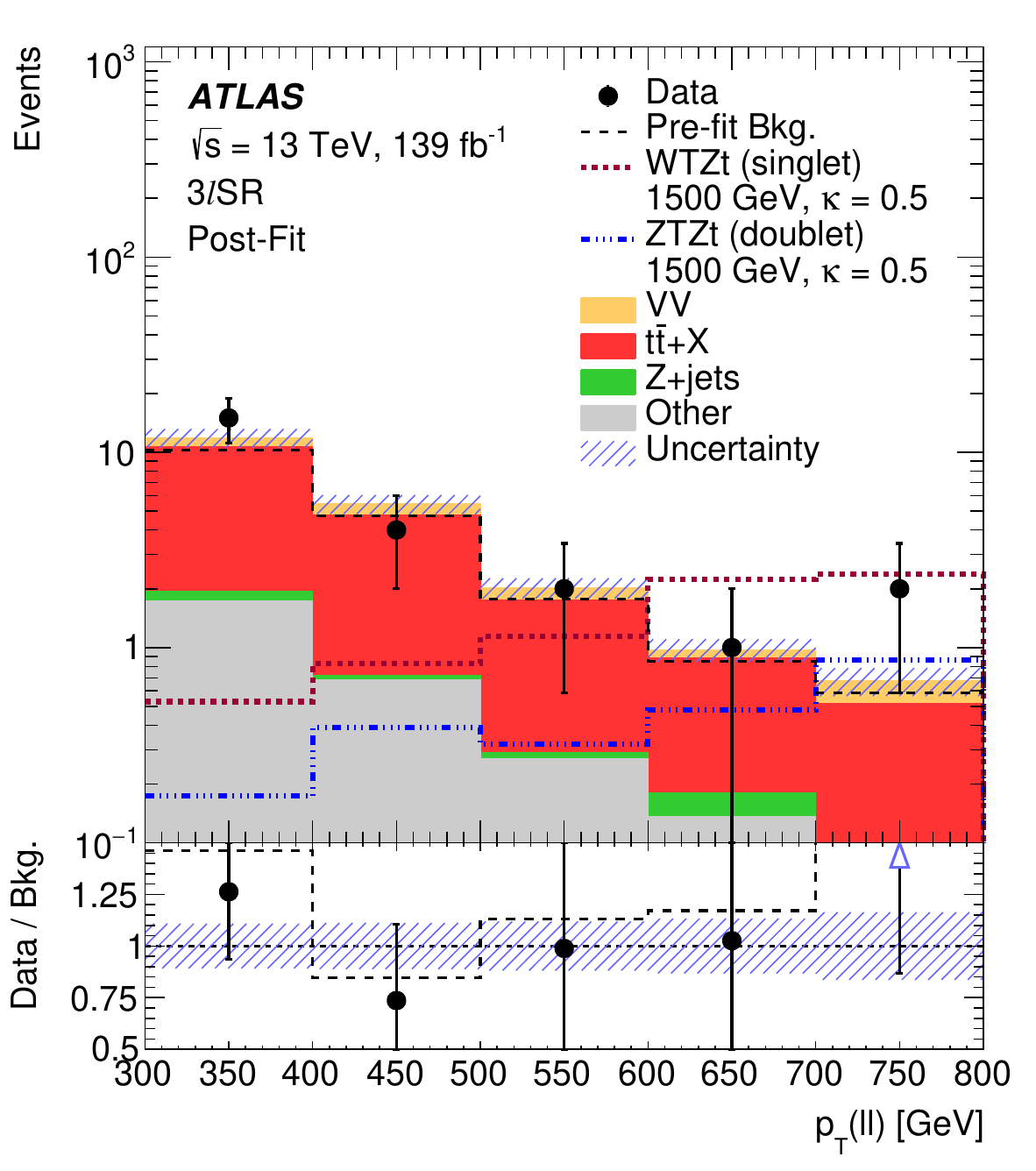}
\label{fig:3l_SR}
}
\caption[]{Distribution of the final discriminant, \ptz, for the \trilep channel in the control regions
\protect\subref{fig:3l_CR1} \trilepreg{VV},
\protect\subref{fig:3l_CR2} \trilepreg{Mixed},
\protect\subref{fig:3l_CR3} \trilepreg{ttX}, the validation region
\protect\subref{fig:3l_VR} \trilepreg{VR}, and the signal region
\protect\subref{fig:3l_SR} \trilepreg{SR}.
The distributions are shown after the background-only fit in the \trilep channel.  The black, dashed line shows the total pre-fit background distribution and the corresponding data-over-background ratio in the bottom panel. The blue arrows in the bottom panel indicate points that are outside the vertical range of the figure.
The overlaid red and blue dotted lines in  \protect\subref{fig:3l_VR}-\protect\subref{fig:3l_SR} show the expected signal contributions in the corresponding regions from $WTZt$ and $ZTZt$ processes  with $M_T = 1.5$~\TeV and $\kappa = 0.5$ in singlet and doublet representations respectively. Expected signal contributions in the control regions are negligible. The last bin in each distribution contains the overflow.
}
\label{fig:3l_CRs}
\end{figure}

\begin{table}[ht!]
\centering
\caption{Observed number of events in data and post-fit number of  background events in the control, validation, and  signal regions for the \trilep channel.  The fit was performed under a background-only hypothesis.
Statistical uncertainties from the limited size of MC samples and systematic uncertainties are added in quadrature. Systematic uncertainties take the correlations among nuisance parameters into account.
The $WTZt$ and $ZTZt$ signal yields correspond to pre-fit yields.
}

\resizebox{\columnwidth}{!}{
\begin{tabular}{|c|r@{\hspace{1.5pt}}c@{\hspace{1.5pt}}l|r@{\hspace{1.5pt}}c@{\hspace{1.5pt}}l|r@{\hspace{1.5pt}}c@{\hspace{1.5pt}}l|r@{\hspace{1.5pt}}c@{\hspace{1.5pt}}l|r@{\hspace{1.5pt}}c@{\hspace{1.5pt}}l|}
\hline

& \multicolumn{3}{c|}{\trilepreg{VV}}
& \multicolumn{3}{c|}{\trilepreg{Mixed}}
& \multicolumn{3}{c|}{\trilepreg{ttX}}
& \multicolumn{3}{c|}{\trilepreg{VR}}
&  \multicolumn{3}{c|}{\trilepreg{SR}}        \\
\hline
$WTZt$ (singlet)  	&  \multirow{2}{*}{3.2}   & \multirow{2}{*}{$\pm$} &  \multirow{2}{*}{{0.4}}
&  \multirow{2}{*}{0.85}  & \multirow{2}{*}{$\pm$} &  \multirow{2}{*}{0.19}
&  \multirow{2}{*}{0.58}  & \multirow{2}{*}{$\pm$} &  \multirow{2}{*}{0.14}
&  \multirow{2}{*}{1.21}  & \multirow{2}{*}{$\pm$} &  \multirow{2}{*}{0.24}
&  \multirow{2}{*}{8.3}   & \multirow{2}{*}{$\pm$} &  \multirow{2}{*}{0.8}     \\
$M_T = 1.5$~TeV, $\kappa=0.5$ & & & & & & & & & & & & & & & \\
\hline
$ZTZt$ (doublet)	&  \multirow{2}{*}{0.84}  & \multirow{2}{*}{$\pm$} &   \multirow{2}{*}{0.24}
&  \multirow{2}{*}{0.68}  &  \multirow{2}{*}{$\pm$} &   \multirow{2}{*}{0.12}
&  \multirow{2}{*}{0.40}  &  \multirow{2}{*}{$\pm$} &   \multirow{2}{*}{0.12}
&   \multirow{2}{*}{2.0}  &  \multirow{2}{*}{$\pm$} &   \multirow{2}{*}{0.8}
&   \multirow{2}{*}{2.04}  &  \multirow{2}{*}{$\pm$} &   \multirow{2}{*}{0.32}     \\
$M_T = 1.5$~TeV, $\kappa=0.5$ & & & & & & & & & & & & & & & \\
\hline\hline
\diboson         	&  {3920} & $\pm$ &  {160}
&  {230}  & $\pm$ &  {40}
&  {21}   & $\pm$ &  {4}
&  {7.1}  & $\pm$ &  {1.5}
&  {2.4}  & $\pm$ &  {0.5}     \\
\ttbar+X    		&  {156}  & $\pm$ &  {27}
&  {245}  & $\pm$ &  {28}
&  {225}  & $\pm$ &  {20}
&  {14.9} & $\pm$ &  {2.5}
&  {15.4} & $\pm$ &  {2.2}     \\
\zjets        		&  {350}  & $\pm$ &  {120}
&  {45}   & $\pm$ &  {13}
&  {3.8}    & $\pm$ &  {1.5}
&  {1.0}  & $\pm$ &  {0.4}
&  {0.31}  & $\pm$ &  {0.11}    \\
\ttbar      		&  {82}   & $\pm$ &  {30}
&  {93}   & $\pm$ &  {16}
&  {5}    & $\pm$ &  {6}
&  {0.9}  & $\pm$ &  {1.1}
&  {0.005}& $\pm$ &  {0.007}   \\
single-top      	&  {63.1} & $\pm$ &  {2.6}
&  {77.9} & $\pm$ &  {1.6}
&  {25.8} & $\pm$ &  {0.8}
&  {2.99} & $\pm$ &  {0.17}
&  {2.88} & $\pm$ &  {0.17}     \\
\triboson        	&  {22.6} & $\pm$ &  {1.2}
&  {1.22} & $\pm$ &  {0.09}
&  {0.087}& $\pm$ &  {0.012}
&  {0.078}& $\pm$ &  {0.015}
&  {0.023}& $\pm$ &  {0.004}   \\
\hline\hline
Total Bkg.      	&  {4594} & $\pm$ &  {70}    %
&  {692}  & $\pm$ &  {26}
&  {281}  & $\pm$ &  {16}
&  {27.0} & $\pm$ &  {3.0}
&  {21.0} & $\pm$ &  {2.2}      \\
\hline\hline
Data             	&  \multicolumn{3}{c|}{{4590}}
&  \multicolumn{3}{c|}{{690}}
&  \multicolumn{3}{c|}{{279}}
&  \multicolumn{3}{c|}{{21}}
&  \multicolumn{3}{c|}{24} \\
\hline
\end{tabular}
}
\label{tab:trilep}
\end{table}


To test whether the observed data is compatible with the background-only hypothesis, tests
are performed with \textsc{RooStats}~\cite{Moneta:2010pm} with statistical models implemented using \textsc{RooFit}~\cite{RooFit} %
and \textsc{HistFactory}~\cite{Cranmer:1456844}.
No significant excess over the background expectation is observed. Hence,
the results were used to set upper limits on the top partner total production cross-section $\sigma(WTZt + ZTZt)$ at 95\% confidence level (CL) using the $\mathrm{CL}_s$ method~\cite{Junk:1999kv,Read:2002hq}. These limits were calculated with the asymptotic approximation~\cite{Cowan:2010js} of the test statistic and validated against the limits obtained using pseudoexperiments for benchmark signal samples with $M_T = 1200,1500,$~and~$2100$~\GeV. For all these benchmarks, the expected and observed limits obtained from pseudoexperiments at 95\% CL agreed with the limits obtained using the asymptotic approximation within 5\%. The largest difference, close to $30\%$, in the results obtained with the two methods was observed in the $-1\sigma$ and $-2\sigma$ bands for the 2100~\GeV mass point.

The overall impact of the systematic uncertainties introduced in  Section~\ref{sec:syst} on the final result is found to be small. Considering only statistical uncertainties, the limits on the total cross-section of the $WTZt$ and $ZTZt$ processes are reduced by less than 14\%.

Given that the signal efficiencies for the $WTZt$ and $ZTZt$ signal modes can be different,
limits were independently calculated for different combinations of top partner mass, coupling, and branching ratios, under the assumption that $\xi_Z = \xi_H$.
The latter is motivated by the Goldstone equivalence theorem~\cite{Goldstone}, which states that the asymptotic branching
ratios of $T$ decaying into $Zt$ and $Ht$ become similar in the large $M_T$ limit under the narrow-width approximation.

To maximize the sensitivity of the search, the results from both channels were statistically combined.
All experimental uncertainties are treated as fully correlated across the channels while the theoretical and modeling uncertainties are taken to be uncorrelated.
The corresponding limits on the signal strength translate into limits on the total cross-section of the $WTZt$ and $ZTZt$ processes
for the corresponding choice of coupling and branching ratio. Limits corresponding to $\kappa = 0.3, 0.5,$~and~0.7 for the singlet and doublet
representations are shown in Figure~\ref{fig:xslims}. The combined expected limit is significantly stronger than limits
obtained independently from the \dilep and \trilep channels. For the singlet representation with $\kappa = 0.5$, top partner masses
less than 1825~\GeV are excluded. The doublet representation only receives contribution from the $ZTZt$ process that has a much smaller cross-section
because of the suppressed matrix element contribution from the gluon splitting into $t\bar{t}$ (Figure~\ref{fig:feyn-WTZt-ZTZt}).
Hence, the doublet exclusion limits are found to be considerably weaker than the singlet limits and none of the considered $T$ masses  are excluded for $\kappa \le 0.5$.

\begin{figure}
\centering
\subfloat[]{
\includegraphics[width=0.46\textwidth]{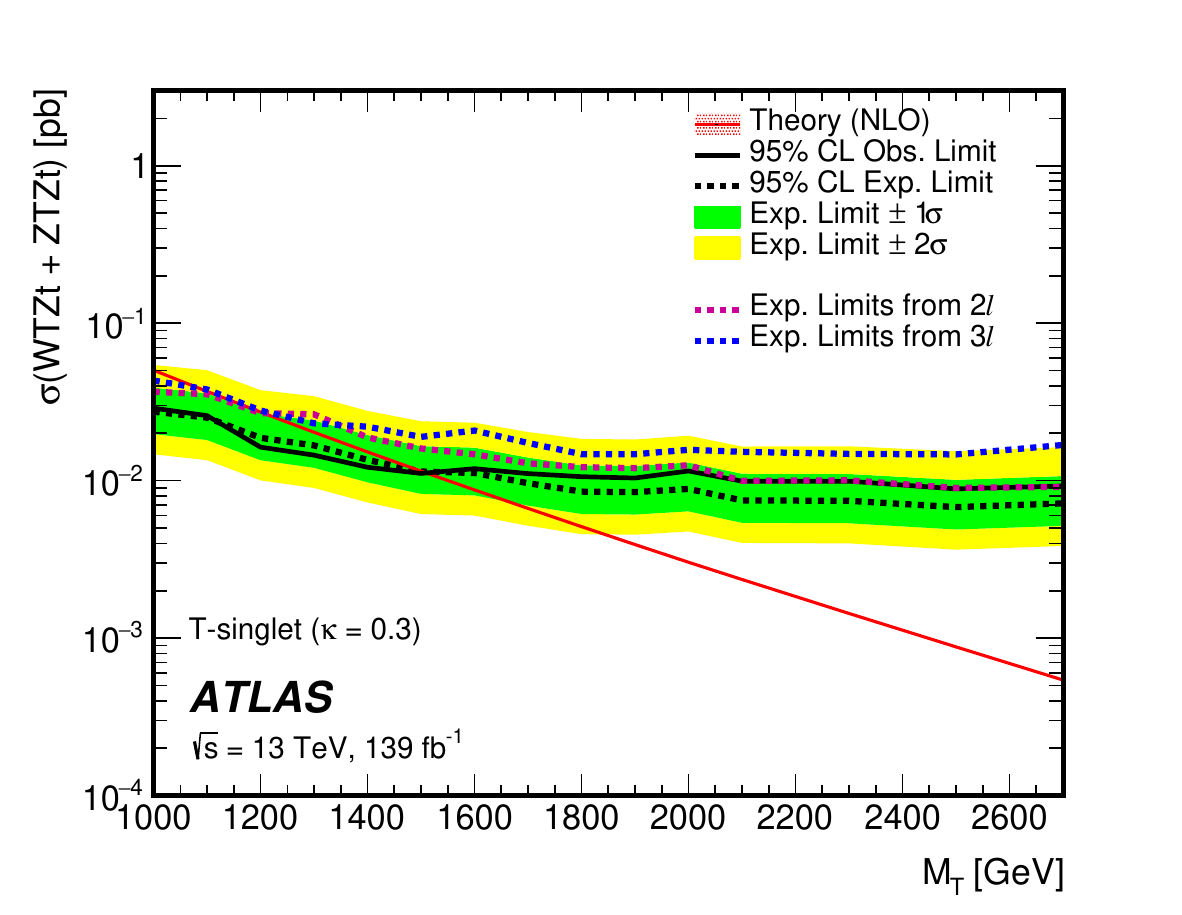}
\label{fig:xslim_singlet_30}
}
\subfloat[]{
\includegraphics[width=0.46\textwidth]{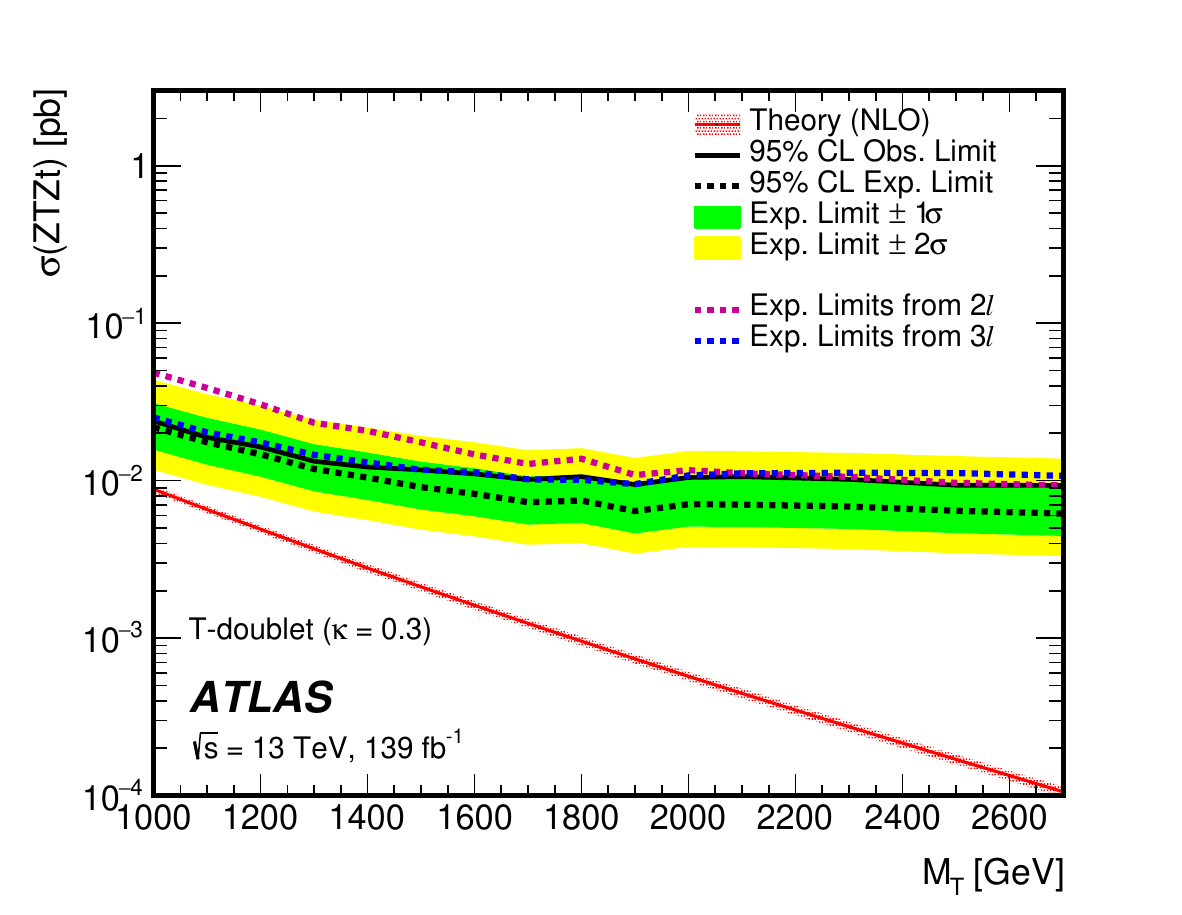}
\label{fig:xslim_doublet_30}
}\\[-2ex]
\subfloat[]{
\includegraphics[width=0.46\textwidth]{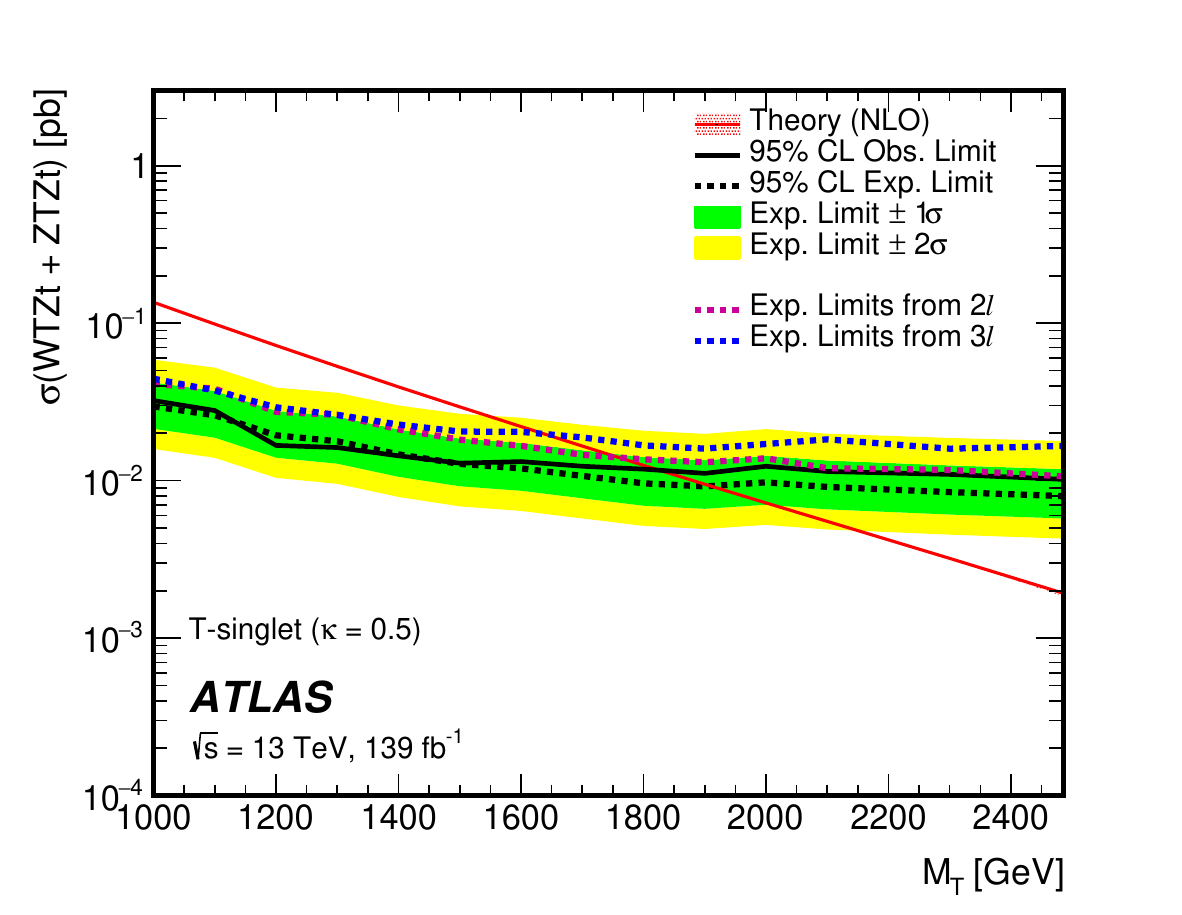}
\label{fig:xslim_singlet_50}
}
\subfloat[]{
\includegraphics[width=0.46\textwidth]{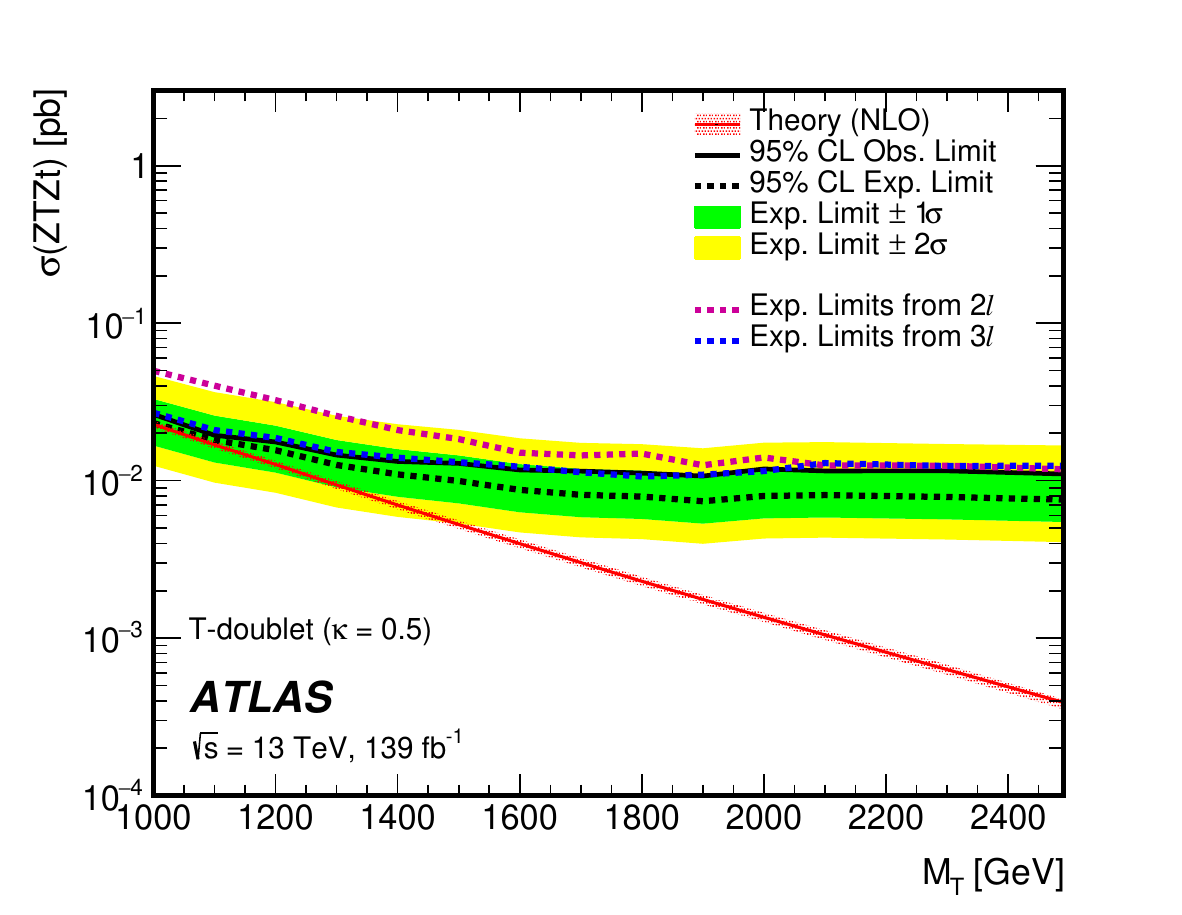}
\label{fig:xslim_doublet_50}
}\\[-2ex]
\subfloat[]{
\includegraphics[width=0.46\textwidth]{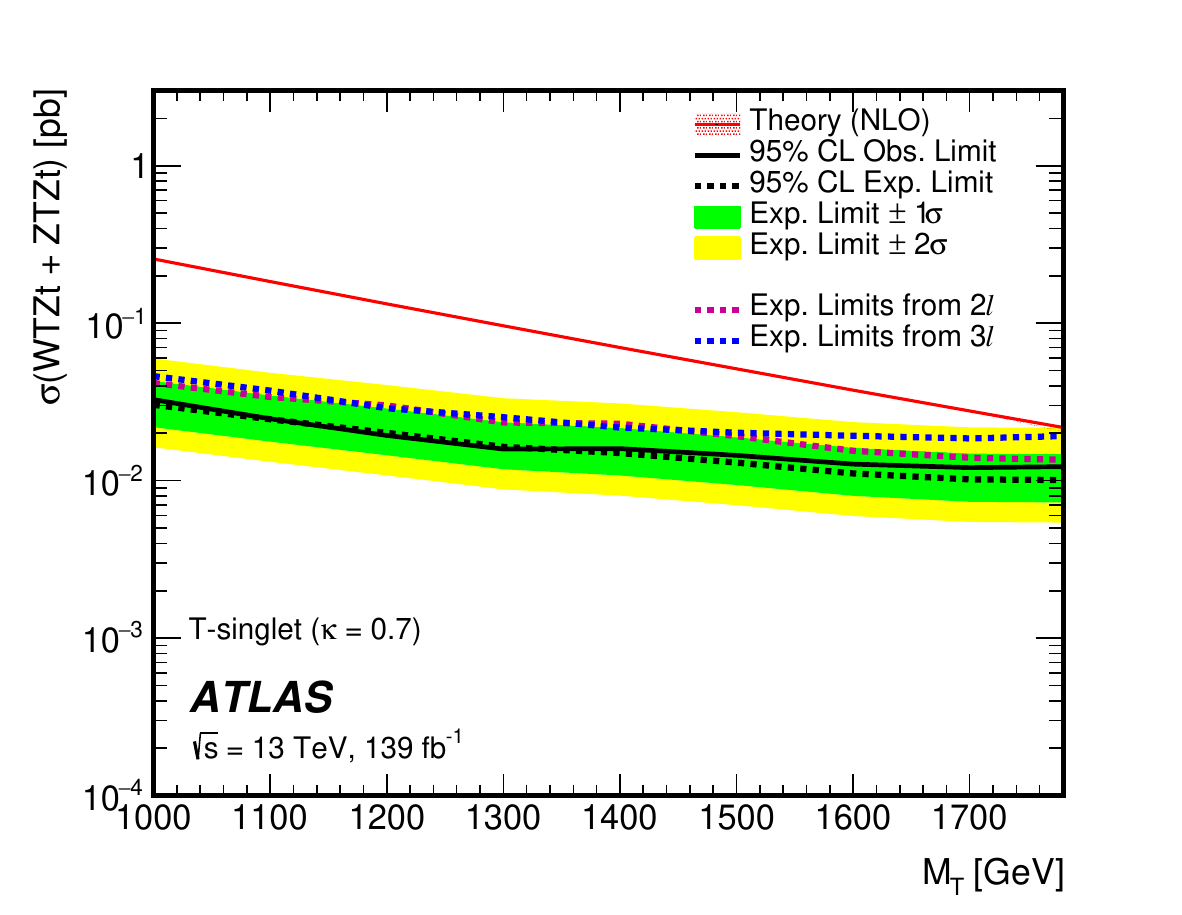}
\label{fig:xslim_singlet_70}
}
\subfloat[]{
\includegraphics[width=0.46\textwidth]{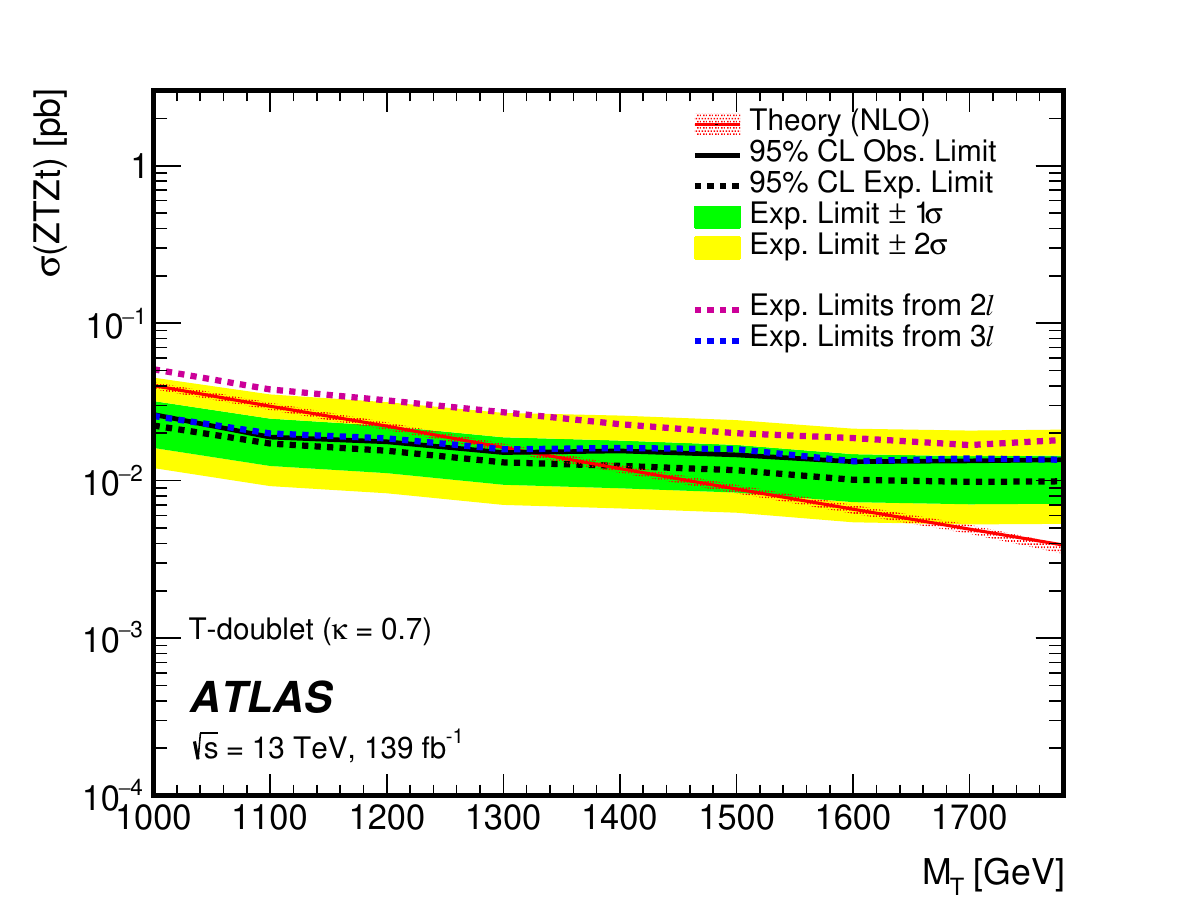}
\label{fig:xslim_doublet_70}
}
\caption[]{Observed (solid black line) and expected (dashed black line) limits calculated at 95\% CL on total cross-section of the $WTZt$ and $ZTZt$ processes as a function of the $T$ mass for a choice of coupling  $\kappa = 0.3$ (top row),  $\kappa = 0.5$ (middle row), and  $\kappa = 0.7$ (bottom row)  for  singlet (left column) and  doublet (right column) representations.

Expected limits calculated at 95\% CL by independently fitting the \dilep and \trilep channels are shown as overlaid blue and red dotted lines respectively.
The green (yellow) band is the 68\% (95\%) confidence interval around the median expected limit.
The solid red line shows the theory prediction for the  NLO cross-section, with the surrounding shaded band representing the corresponding uncertainty.
The mass range in these figures is restricted to ensure that the relative decay width $\frac{\Gamma_T}{M_T}$ is limited to less than 50\%.
}
\label{fig:xslims}
\end{figure}

The cross-section limits calculated for different choices of coupling can be reinterpreted as limits on the plane of $(M_T, \kappa)$, as shown in Figures~\ref{fig:kappalim_singlet}~and~\ref{fig:kappalim_doublet} for the singlet and doublet representations only.
Following the prescription of Ref.~\cite{Panizzi-LW-3}, this interpretation is only applied for the parametric subspace where the top partner's relative decay width, $\frac{\Gamma_T}{M_T}$ is smaller than 50\%.
The correction factors for finite width and non-resonant contributions~\cite{interpretation, vlqnonres} are also valid for the same parametric subspace. Hence, the limits in Figure~\ref{fig:kappalims} beyond $\frac{\Gamma_T}{M_T} > 50\%$  are omitted for phenomenological consistency.

\begin{figure}
\centering
\subfloat[]{
\includegraphics[width=0.48\textwidth]{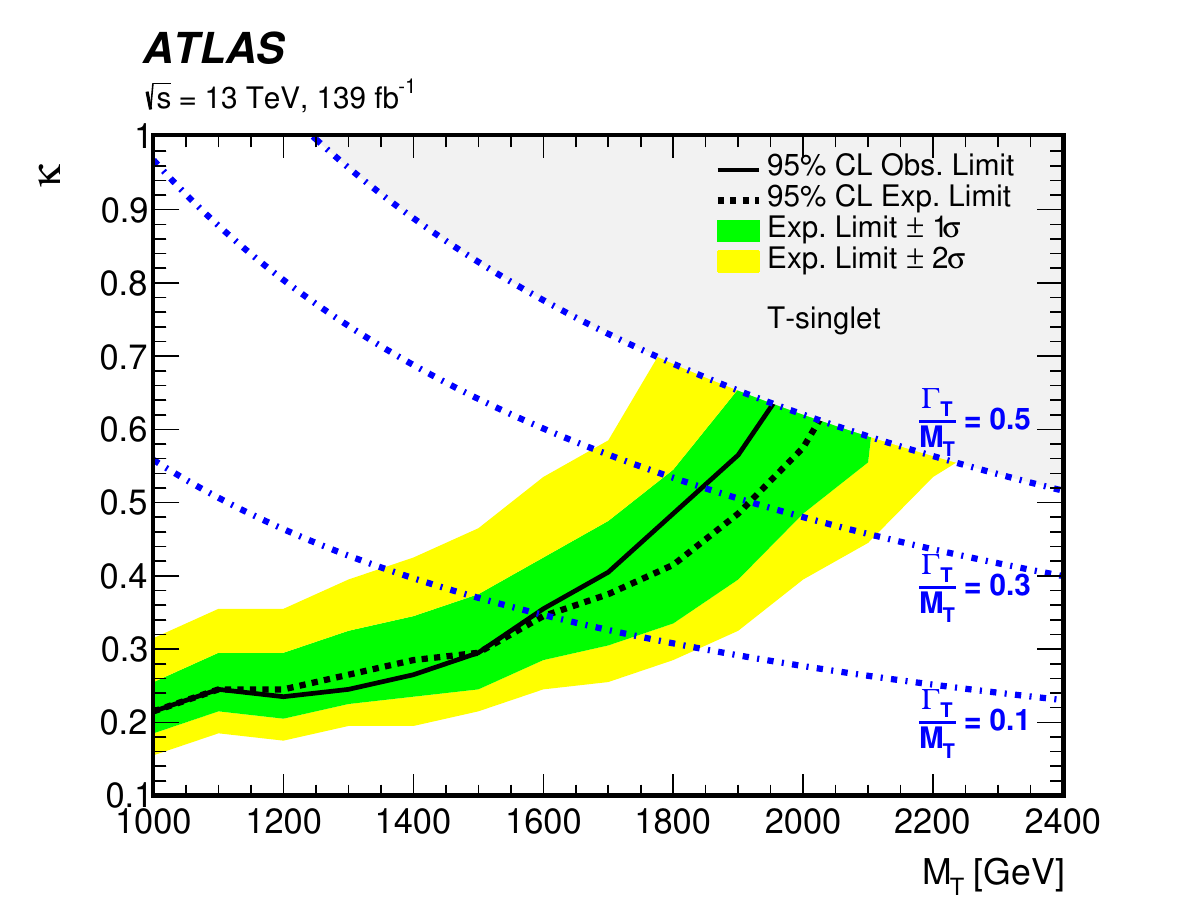}
\label{fig:kappalim_singlet}
}
\subfloat[]{
\includegraphics[width=0.48\textwidth]{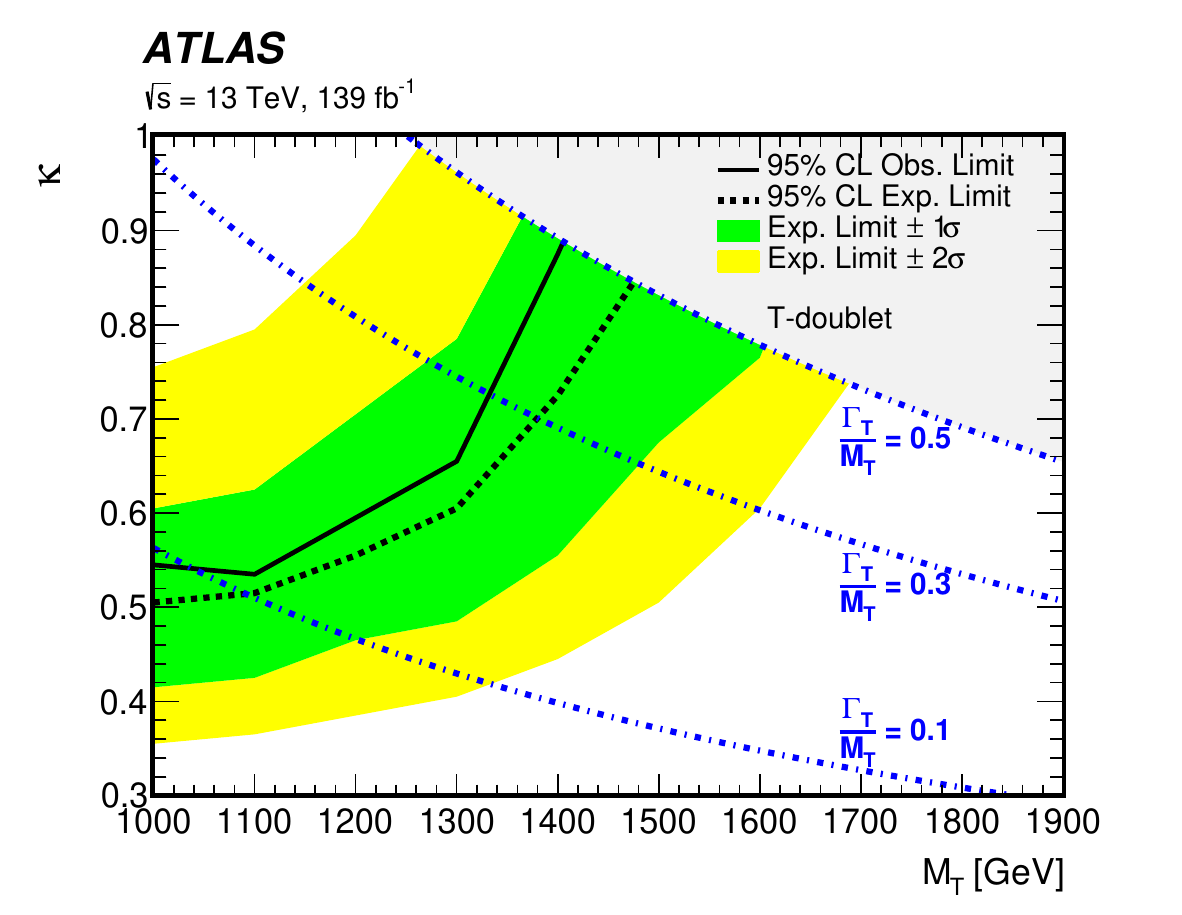}
\label{fig:kappalim_doublet}
}
\caption[]{Observed (solid black line) and expected (dashed black line) limits calculated at 95\% CL on the top partner coupling as a function of the $T$ mass for \subref{fig:kappalim_singlet} singlet and \subref{fig:kappalim_doublet} doublet representations.
The shaded regions represent the parametric space with $\frac{\Gamma_T}{M_T} > 50\%$.
The green (yellow) band is the 68\% (95\%) confidence interval around the median expected limit.
The dotted lines represent contours of equal values in $\frac{\Gamma_T}{M_T}$ .
}
\label{fig:kappalims}
\end{figure}

To further generalize the interpretation of the search results for any possible combination of top partner branching ratios, the semi-analytical interpretation strategy presented in Ref.~\cite{interpretation} was adopted in Figure~\ref{fig:newlims} to represent the excluded top partner mass as a function of its relative decay width and $\xi_W$, the asymptotic branching ratio for the $T \to Wb$ decay mode. The $\xi$ parameters for the other two decay modes are determined by the constraint $\xi_Z = \xi_H = \frac{1-\xi_W}{2}$.

\begin{figure}
\centering
\subfloat[]{
\includegraphics[width=0.48\textwidth]{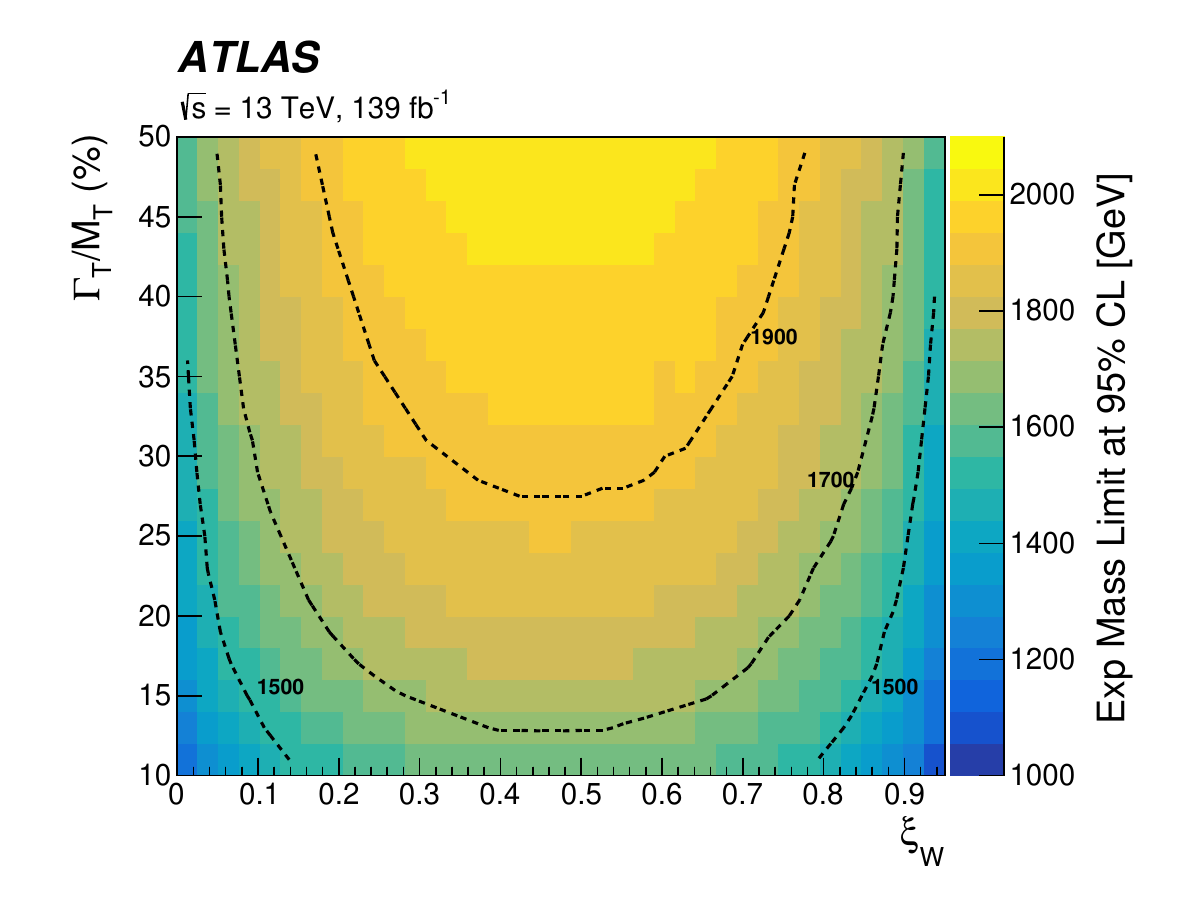}
\label{fig:newlim_exp}
}
\subfloat[]{
\includegraphics[width=0.48\textwidth]{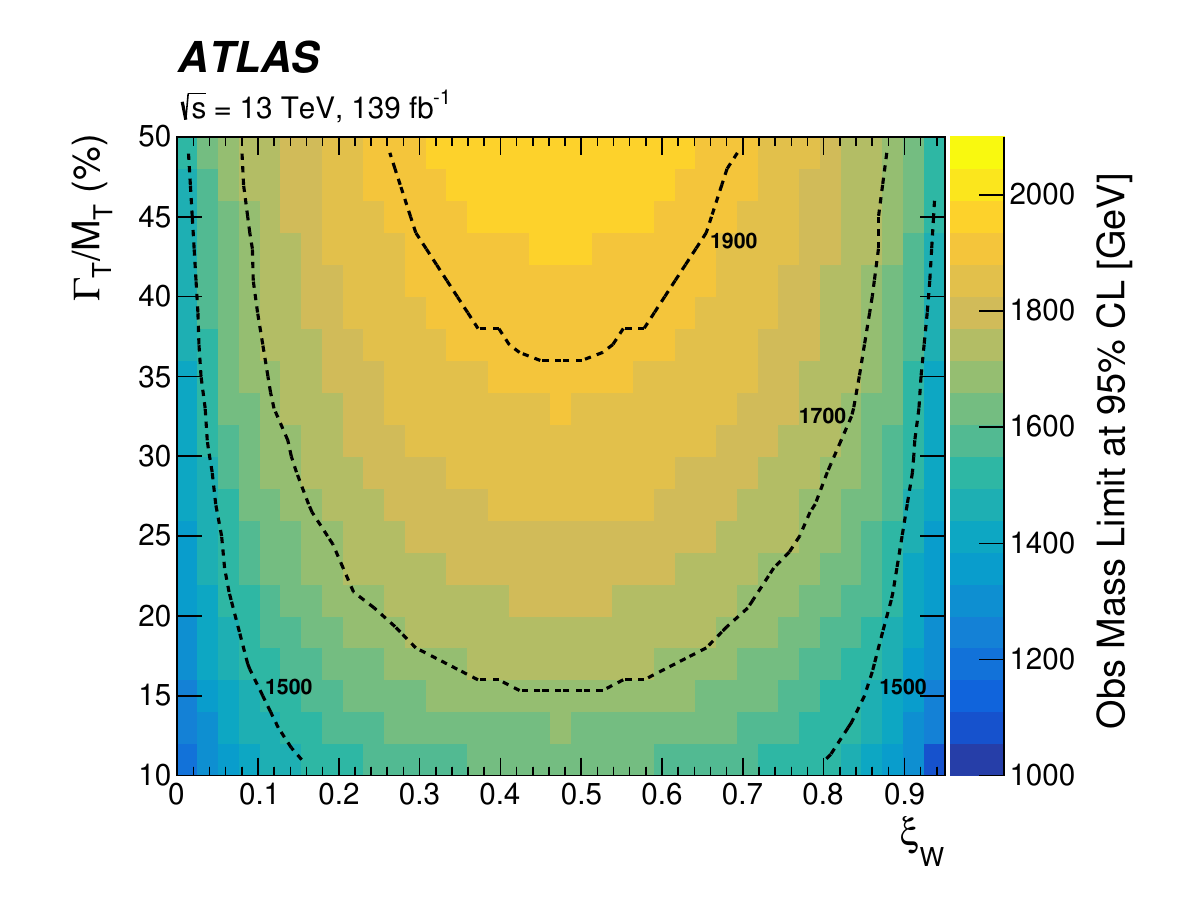}
\label{fig:newlim_obs}
}
\caption[]{\subref{fig:newlim_exp} Expected and \subref{fig:newlim_obs} observed upper limits calculated at 95\% CL on the excluded top partner mass as a function of its relative decay width $(\frac{\Gamma_T}{M_T})$ and the relative coupling parameter $\xi_W$ with the assumption of $\xi_Z = \xi_H = \frac{1-\xi_W}{2}$. The dotted contour lines represent exclusion limits of equal mass in units of \GeV.
}
\label{fig:newlims}
\end{figure}


%

%

%
%
\FloatBarrier

\section{Conclusion}
\label{sec:conclusion}

A search for the  production of a single vector-like quark  $T$ with electric charge $\frac{2}{3}e$
and mass greater than 1~TeV
in a final state with  a leptonically decaying $Z$ boson and a top quark is presented.
Two orthogonal channels with two or at least three leptons are separately
optimized and their results are statistically combined to obtain the  final result.
No  significant excess over the background expectation is observed, and therefore 95\% CL upper limits on the inclusive single-$T$ cross-section are derived for different choices
of $T$ mass and couplings. These results are interpreted in terms of limits on the $T$ mass and coupling for different electroweak
representations and generalized branching ratio scenarios.
For the singlet representation,   $\kappa$ values are excluded between 0.22 and 0.64 for masses of $T$ between 1000 and 1975~\GeV.
The exclusion range for the doublet representation spans $\kappa$ values between 0.54 and 0.88 for masses of $T$ between 1000 and 1425~\GeV.
These limits significantly improve on the previous limits obtained from the 2015--16 data sample. The strongest exclusion is observed for singlet representation with $\xi_W \approx 0.5$ where masses up to 1975~\GeV are excluded at relative decay width of $\frac{\Gamma_T}{M_T} = 0.5$ for the top partner.
The exclusion of singlet $T$ production with mass and coupling of 1.8 (1.6)~TeV and 0.70 (0.35), corresponding to a decay width of 30\% (10\%), provides the strongest limits from the LHC in these parameter regions.

%


%

%

%
%
\section*{Acknowledgments}
%

%
%

%
%

We thank CERN for the very successful operation of the LHC, as well as the
support staff from our institutions without whom ATLAS could not be
operated efficiently.

We acknowledge the support of
ANPCyT, Argentina;
YerPhI, Armenia;
ARC, Australia;
BMWFW and FWF, Austria;
ANAS, Azerbaijan;
CNPq and FAPESP, Brazil;
NSERC, NRC and CFI, Canada;
CERN;
ANID, Chile;
CAS, MOST and NSFC, China;
Minciencias, Colombia;
MEYS CR, Czech Republic;
DNRF and DNSRC, Denmark;
IN2P3-CNRS and CEA-DRF/IRFU, France;
SRNSFG, Georgia;
BMBF, HGF and MPG, Germany;
GSRI, Greece;
RGC and Hong Kong SAR, China;
ISF and Benoziyo Center, Israel;
INFN, Italy;
MEXT and JSPS, Japan;
CNRST, Morocco;
NWO, Netherlands;
RCN, Norway;
MEiN, Poland;
FCT, Portugal;
MNE/IFA, Romania;
MESTD, Serbia;
MSSR, Slovakia;
ARRS and MIZ\v{S}, Slovenia;
DSI/NRF, South Africa;
MICINN, Spain;
SRC and Wallenberg Foundation, Sweden;
SERI, SNSF and Cantons of Bern and Geneva, Switzerland;
MOST, Taiwan;
TENMAK, T\"urkiye;
STFC, United Kingdom;
DOE and NSF, United States of America.
In addition, individual groups and members have received support from
BCKDF, CANARIE, Compute Canada and CRC, Canada;
PRIMUS 21/SCI/017 and UNCE SCI/013, Czech Republic;
COST, ERC, ERDF, Horizon 2020 and Marie Sk{\l}odowska-Curie Actions, European Union;
Investissements d'Avenir Labex, Investissements d'Avenir Idex and ANR, France;
DFG and AvH Foundation, Germany;
Herakleitos, Thales and Aristeia programmes co-financed by EU-ESF and the Greek NSRF, Greece;
BSF-NSF and MINERVA, Israel;
Norwegian Financial Mechanism 2014-2021, Norway;
NCN and NAWA, Poland;
La Caixa Banking Foundation, CERCA Programme Generalitat de Catalunya and PROMETEO and GenT Programmes Generalitat Valenciana, Spain;
G\"{o}ran Gustafssons Stiftelse, Sweden;
The Royal Society and Leverhulme Trust, United Kingdom.

The crucial computing support from all WLCG partners is acknowledged gratefully, in particular from CERN, the ATLAS Tier-1 facilities at TRIUMF (Canada), NDGF (Denmark, Norway, Sweden), CC-IN2P3 (France), KIT/GridKA (Germany), INFN-CNAF (Italy), NL-T1 (Netherlands), PIC (Spain), ASGC (Taiwan), RAL (UK) and BNL (USA), the Tier-2 facilities worldwide and large non-WLCG resource providers. Major contributors of computing resources are listed in Ref.~\cite{ATL-SOFT-PUB-2021-003}.

%
%


%
%
%
%
%
%
%

%
\clearpage
\printbibliography
\clearpage
\input{atlas_authlist}
%

%
%
%

%

%
%
%
%

\end{document}

%% file: atlas_authlist.tex
 
\begin{flushleft}
\hypersetup{urlcolor=black}
{\Large The ATLAS Collaboration}

\bigskip

\AtlasOrcid[0000-0002-6665-4934]{G.~Aad}$^\textrm{\scriptsize 102}$,
\AtlasOrcid[0000-0002-5888-2734]{B.~Abbott}$^\textrm{\scriptsize 120}$,
\AtlasOrcid[0000-0002-1002-1652]{K.~Abeling}$^\textrm{\scriptsize 55}$,
\AtlasOrcid[0000-0001-5763-2760]{N.J.~Abicht}$^\textrm{\scriptsize 49}$,
\AtlasOrcid[0000-0002-8496-9294]{S.H.~Abidi}$^\textrm{\scriptsize 29}$,
\AtlasOrcid[0000-0002-9987-2292]{A.~Aboulhorma}$^\textrm{\scriptsize 35e}$,
\AtlasOrcid[0000-0001-5329-6640]{H.~Abramowicz}$^\textrm{\scriptsize 151}$,
\AtlasOrcid[0000-0002-1599-2896]{H.~Abreu}$^\textrm{\scriptsize 150}$,
\AtlasOrcid[0000-0003-0403-3697]{Y.~Abulaiti}$^\textrm{\scriptsize 117}$,
\AtlasOrcid[0000-0003-0762-7204]{A.C.~Abusleme~Hoffman}$^\textrm{\scriptsize 137a}$
% ...
\bigskip
\\

% ...
$^{29}$Physics Department, Brookhaven National Laboratory, Upton NY; United States of America.\\
% ...
$^{35}$$^{(a)}$Facult\'e des Sciences Ain Chock, R\'eseau Universitaire de Physique des Hautes Energies - Universit\'e Hassan II, Casablanca;$^{(b)}$Facult\'{e} des Sciences, Universit\'{e} Ibn-Tofail, K\'{e}nitra;$^{(c)}$Facult\'e des Sciences Semlalia, Universit\'e Cadi Ayyad, LPHEA-Marrakech;$^{(d)}$LPMR, Facult\'e des Sciences, Universit\'e Mohamed Premier, Oujda;$^{(e)}$Facult\'e des sciences, Universit\'e Mohammed V, Rabat;$^{(f)}$Institute of Applied Physics, Mohammed VI Polytechnic University, Ben Guerir; Morocco.\\
% ...
$^{49}$Fakult\"{a}t Physik , Technische Universit{\"a}t Dortmund, Dortmund; Germany.\\
% ...
$^{55}$II. Physikalisches Institut, Georg-August-Universit\"{a}t G\"ottingen, G\"ottingen; Germany.\\
% ...
$^{102}$CPPM, Aix-Marseille Universit\'e, CNRS/IN2P3, Marseille; France.\\
% ...
$^{117}$Department of Physics, New York University, New York NY; United States of America.\\
% ...
$^{120}$Homer L. Dodge Department of Physics and Astronomy, University of Oklahoma, Norman OK; United States of America.\\
% ...
$^{137}$$^{(a)}$Departamento de F\'isica, Pontificia Universidad Cat\'olica de Chile, Santiago;$^{(b)}$Millennium Institute for Subatomic physics at high energy frontier (SAPHIR), Santiago;$^{(c)}$Instituto de Investigaci\'on Multidisciplinario en Ciencia y Tecnolog\'ia, y Departamento de F\'isica, Universidad de La Serena;$^{(d)}$Universidad Andres Bello, Department of Physics, Santiago;$^{(e)}$Instituto de Alta Investigaci\'on, Universidad de Tarapac\'a, Arica;$^{(f)}$Departamento de F\'isica, Universidad T\'ecnica Federico Santa Mar\'ia, Valpara\'iso; Chile.\\
% ...
$^{150}$Department of Physics, Technion, Israel Institute of Technology, Haifa; Israel.\\
$^{151}$Raymond and Beverly Sackler School of Physics and Astronomy, Tel Aviv University, Tel Aviv; Israel.\\
% ...
$^{a}$ Also Affiliated with an institute covered by a cooperation agreement with CERN.\\
$^{b}$ Also at An-Najah National University, Nablus; Palestine.\\
$^{c}$ Also at Borough of Manhattan Community College, City University of New York, New York NY; United States of America.\\
$^{d}$ Also at Center for High Energy Physics, Peking University; China.\\
$^{e}$ Also at Center for Interdisciplinary Research and Innovation (CIRI-AUTH), Thessaloniki; Greece.\\
$^{f}$ Also at Centro Studi e Ricerche Enrico Fermi; Italy.\\
% ...

\end{flushleft}
